\documentclass[a4paper,12pt]{article}

\usepackage[a4paper,left=80pt,right=80pt,top=90pt,bottom=98pt]{geometry}
\usepackage[latin1]{inputenc}
\usepackage{dsfont}
\usepackage{amsfonts,amsmath,amssymb}
\usepackage{amsthm}
\usepackage{graphicx}
\usepackage[small,bf]{caption}
\usepackage{subcaption}
\usepackage{booktabs}
\usepackage[numbers, sort]{natbib}
\allowdisplaybreaks[1]
\usepackage{color}
\usepackage{ulem}
\usepackage{tikz}
\usepackage{multirow}
\usetikzlibrary{arrows}
\usepackage{xspace}
\def\bal#1\eal{\begin{align}#1\end{align}}
\newcommand{\be}{\begin{equation}}
\newcommand{\ee}{\end{equation}}
\newcommand{\bea}{\begin{eqnarray}}
\newcommand{\eea}{\end{eqnarray}}
\newcommand{\besub}{\begin{subequations}}
\newcommand{\eesub}{\end{subequations}}
\newcommand{\ba}{\begin{array}}
\newcommand{\ea}{\end{array}}
\newcommand{\bi}{\begin{itemize}}
\newcommand{\ei}{\end{itemize}}

\begin{document}

\begin{titlepage}

%\flushright{HIP-2019-XX/TH}

\vspace*{1cm}

\begin{center}
{\Large 
{\bf
Real Scalar Dark Matter: Relativistic Treatment 
%and Chemical Potential Effects
}}
\\
[2cm]

{
{\bf
Giorgio Arcadi$^{1,2,3}$, Oleg Lebedev$^4$, Stefan Pokorski$^5$, Takashi Toma$^6$
 }}

\vspace{1.5cm}
 {\it {\small $^1$Max-Planck-Institut  fur  Kernphysik, Saupfercheckweg  1,  69117  Heidelberg,  Germany\\
  
  \vspace{0.2cm}
  $^2$Dipartimento di Matematica e Fisica, Universit\`a di Roma
3, Via della Vasca Navale 84, 00146, Roma, Italy\\

\vspace{0.2cm}
$^3$INFN Sezione Roma 3, Italy\\

 \vspace{0.3cm}
 $^4$Department of Physics, University of Helsinki, Gustaf H\"allstr\"omin katu 2a, Helsinki, Finland

\vspace{0.3cm}
$^5$Institute of Theoretical Physics, Faculty of Physics, University of Warsaw, ul. Pasteura 5, PL-02-093 Warsaw, Poland

 \vspace{0.3cm}
 $^6$Department of Physics, McGill University,\\
 3600 Rue University, Montr\'{e}al, Qu\'{e}bec H3A 2T8, Canada\\ }}

\end{center}
%\addtocounter{footnote}{-4}

\vspace*{0.5cm}

\vspace*{0.4cm}

\vspace*{1.2cm}

\begin{abstract}
\noindent
 A stable real scalar  provides one of the simplest possibilities  to account for dark matter. We consider the regime where its coupling to the Standard Model fields is negligibly small. Due to self--coupling, the scalar field can reach thermal or at least kinetic equilibrium, in which case the system is characterized by its temperature and effective chemical potential. 
 We perform a fully relativistic analysis of dark matter evolution, thermalization conditions  and  different freeze--out regimes, including the chemical potential effects.
  To this end, we derive a relativistic Bose--Einstein analog of the Gelmini--Gondolo formula for a thermal averaged cross section. Finally, we perform a comprehensive parameter space analysis to determine regions consistent with observational constraints. Dark matter can be both warm and cold in this model.
    \end{abstract}

%\today

\end{titlepage}
\newpage

\tableofcontents

%=========================================================================
%=========================================================================
\section{Introduction}
%=========================================================================
%=========================================================================

The nature of dark matter (DM) remains one of the outstanding questions of modern physics. The popular WIMP (weakly interacting massive particle) paradigm currently finds itself under pressure due to ever improving bounds on dark matter interaction with nucleons \cite{Aprile:2018dbl}. This motivates one to explore alternative scenarios in which the dark matter relic  density is not related to the annihilation into Standard Model (SM) particles. In this work, we consider in detail one such example. In this case, the DM relic density is set by either the dark sector thermodynamics or postinflationary initial conditions.

We study the simplest dark sector consisting of a single real scalar field  $S$ with the potential       \cite{Silveira:1985rk}
 \begin{equation}
 V= {m^2\over 2} S^2 + {\lambda\over 4!} S^4 \;,
 \end{equation}
 where the self--coupling $\lambda$ is large enough for the $S$ quanta to reach   kinetic equilibrium at least. 
We  assume that the dark matter coupling to the Standard Model, $\lambda_{HS} H^\dagger H S^2$,
is negligibly small, $\lambda_{HS} < 10^{-13}$. In this case, the observable and dark sectors do not equilibrate, DM freeze--in production  \cite{Hall:2009bx}   is suppressed 
 and the DM evolution is determined entirely by its self--coupling \cite{Carlson:1992fn}.

Although we focus on just one model of dark matter, the novelty of our approach is that we account for 
all the stages in the  (thermal)  dynamics,  which requires a relativistic treatment with the Bose--Einstein distribution function.  In particular, we
derive  relativistic expressions for the number conserving and number changing reaction rates generalizing the well known Gelmini--Gondolo result \cite{Gondolo:1990dk}, and solve the corresponding Boltzmann equation coupled with the entropy conservation condition. This allows us to study the different freeze--out regimes, including the relativistic one,   as well as trace the transition from the pre-freeze--out epoch to post-freeze--out by deriving the chemical potential evolution.  We consider both fully thermalized dark matter and DM in kinetic equilibrium. Finally, we delineate parameter 
space consistent with all the observational constraints, for which 
a relativistic treatment    is essential.

Regarding observational prospects, dark matter detection would be very  challenging due its negligible interaction with the visible sector.
 One 
may potentially observe effects due to its self--interaction, but even that is not guaranteed because $\lambda$ is allowed to be very small and still
consistent with the relic density constraints. 

 Previous analyses of this  or closely related models include Refs.\;\cite{McDonald:1993ex}-\cite{Binder:2017rgn}
 where $S$ is treated as a WIMP--like particle, and non--relativistic studies of feebly interacting $S$ 
 \cite{Bernal:2015bla}-\cite{Fairbairn:2018bsw}. 
 We go beyond these results in that we offer a fully relativistic treatment and also  include the chemical potential.
 Relativistic effects in the Boltzmann equation within other contexts have been considered, for example,
  in  \cite{Dolgov:1992wf}-\cite{Olechowski:2018xxg}. These analyses are not however applicable to the model at hand.
 
The paper is organized as follows: after providing an inflationary motivation for our study, we derive the expressions for the relativistic reactions rates. We then apply them
to derive the thermalization and freeze--out conditions,  and delineate parameter space consistent with the observed DM relic abundance.

\section{Inflationary motivation}

 In this section, we motivate the setting for our study. Our main assumptions are that the interaction between the SM fields and dark matter (DM) is negligible, and that DM reaches some degree of thermal equilibrium, be it kinetic or chemical one.   
 We argue  that the temperatures of the observable and dark sectors can differ dramatically, and that the latter may also be characterized by   effective chemical potential.

 \subsection{Initial conditions for dark matter  evolution} 
 
A bosonic system in thermal or kinetic  equilibrium is described by its temperature $T$ and effective chemical potential $\mu$
via the momentum  distribution
\begin{equation}
 f(p) ={ 1  \over e^{E-\mu \over T} -1  }   \;,
 \end{equation}
 where $E= \sqrt{m^2+ \vec{p}^2}$. Here the chemical potential is not necessarily associated with the conserved Noether current, instead it reflects approximate conservation of a particle number in a given regime. The above distribution applies also to bosonic systems in an expanding FRW Universe \cite{Bernstein:1985th,Bernstein:1988bw,Kolb:1990vq}.

 In general, the dark matter temperature $T$ can be very different from that of the observable sector $T_{\rm SM}$, as long
as the interaction between the two is very weak \cite{Carlson:1992fn}. Furthermore, if the number changing processes in the dark sector are inefficient, dark matter can be assigned effective chemical potential  $\mu_{\rm  }$ which determines its number density at a given temperature,
\begin{equation}
 n_{\rm  } ={1\over 2 \pi^2} \int_{m_{\rm  }}^\infty { (E^2-m_{\rm  }^2)^{1/2} \over e^{E-\mu_{\rm  } \over T'} -1  } E dE \;,
 \end{equation}
where $\mu_{\rm } <m_{\rm  }$.
Even in the simplest case of real scalar dark matter, both $T$ and $\mu_{\rm  }$ are in general necessary to describe its state. These determine the initial conditions for the evolution of the system from the relativistic regime to freeze--out and its eventual state we observe today.

\subsection{Example}

Dark matter and observable matter may be produced by very different mechanisms resulting in  different thermodynamic properties of the two sectors. In particular, their temperatures may differ by orders of magnitude and the hidden sector can be endowed with chemical potential if the number changing processes are inefficient.

Consider a simple possibility that the observed matter and dark matter are produced 
after inflation via direct interaction with the inflaton $\phi$. Suppose that the leading 
interaction terms are 
\begin{eqnarray}
&& {V}_{\rm \phi h}= \sigma  \phi h^2 \;,\\
&& {V}_{\rm \phi S}= {1\over 2} \lambda_{ \phi S}  \phi^2  S^2\;, 
\end{eqnarray} 
where the unitary gauge for the Higgs field  $h$ has been assumed.  For the purpose of illustration, let us also choose the simplest inflaton potential, 
\begin{equation}
   V_{\rm inf}= {1\over 2} m_\phi^2 \phi^2 \;,
 \end{equation} 
 with $m_\phi \sim 10^{13} \;{\rm GeV} \gg m$.    
  The matter production mechanisms depend on the balance between $\sigma$ and
$\lambda_{ \phi S}  \phi$. For relatively small $\sigma$, SM matter is produced at late times via perturbative decay of the inflaton, while dark matter can be produced efficiently  right after inflation through parametric resonance.

Let us consider these mechanisms in more detail.  During the preheating stage, the inflaton oscillates 
with a decreasing amplitude as $\phi(t) \simeq \phi_0/ m_\phi t \;  \sin m_\phi t$, where $\phi_0$ is close to the Planck scale. This creates an oscillating mass term for $S$ and leads to efficient dark matter  production as long as $ q=\lambda_{ \phi S}  \phi^2/m_\phi^2 \gg 1 $. When $q$ reaches 1, the resonance stops and
the subsequent DM evolution is determined primarily by its self--coupling $\lambda$. 
If $\lambda$ is small enough,   the total number of the DM quanta remains approximately constant.

For small $ \lambda_{ \phi S} \lesssim 10^{-8}$, 
 the resonance is efficient for a short time of order 10$m_\phi^{-1}$ and produces relativistic DM quanta with typical momentum $k \sim m_\phi$ (see \cite{Enqvist:2016mqj} for a recent study). It converts only a small fraction of the inflaton energy into radiation.
 The corresponding DM number density is conveniently parametrized by 
 \begin{equation}
  n (a=1) = c \; m_\phi^3 \;, 
 \end{equation} 
  where $c$ depends on the strength of the resonance $q$ and $a$ is the FRW scale factor which we set to one at this stage. 
   The energy of the DM quanta  and their  density evolve in time  as
 \begin{equation}
 E(a) = E(1)/a ~~,~~n(a) = n(1)/a^3 \;.
 \end{equation} 
 As long as  the energy density of the Universe is dominated by the inflaton, $a \propto t^{2/3}$.

Radiation domination sets in when perturbative inflaton decay $\phi \rightarrow hh$  rate becomes comparable to the Hubble rate,
\begin{equation}
\Gamma= {\sigma^2 \over 8 \pi m_\phi} \sim H \simeq {m_\phi \phi_{\rm end} \over \sqrt{6} M_{\rm Pl} a^{3/2}} \;,
\end{equation}
 where $\phi_{\rm end}$ is the inflaton amplitude at the end of the resonance. This equation can be solved for $a$ which plays the role of the time variable.
The resulting reheating temperature is found via the usual relation
\begin{equation}
 T_R \sim 10^{-1}\sqrt{\Gamma M_{\rm Pl}} \;,
\end{equation}
which assumes that nearly all of the inflaton energy converts into Standard Model radiation.\footnote{Here we neglect the $h$ production via tachyonic resonance. This is justified for  $\sigma \phi / m_\phi^2  <  1$.} 
 At this point, the ratio between the average energy of the DM quanta  $E(a)\simeq m_\phi/a$ and $T_R$ is given by 
 \begin{equation}
 {E(a) \over T_R} \sim \left(    {m_\phi \sigma^2 M_{\rm Pl} \over \phi_{\rm end}^4 }  \right)^{1/6} \ll 1
\end{equation}
for $\sigma \ll M_{\rm Pl}$.
This ratio stays constant in time since both quantities scale as radiation. Thus, it determines the ratio between the hidden sector temperature $T$ and that of the observable sector $T_{\rm SM}=T_R$, once dark matter reaches kinetic  equilibrium through   self--interaction,
\begin{equation}
 {E (a) \over T_R } \sim \ {T \over T_{\rm SM} }  \ll 1\;.
\end{equation}
While the temperature is fixed by the average energy of the quanta, the number density is an independent quantity determined by the production mechanism.
It is instructive 
to  compare the DM number density  $n(a)$  to the equilibrium density at temperature  $T(a)$ with zero chemical potential, $n(T) \simeq \zeta(3)/\pi^2 \; T^{ 3}$,
\begin{equation}
{n(a)  \over n (T(a))} \sim {c \pi^2 \over \zeta(3)} \;.
\end{equation}
This quantity  also stays constant in time if the number changing interactions are suppressed. 
Its magnitude  depends strongly on the efficiency of   parametric resonance. If the resonance is weak, $ \lambda_{ \phi S} < 10^{-8}$, one finds $c\ll 0.1$ and ${n(a)  / n (T(a))}\ll 1$.\footnote{We thank Stanislav Rusak for verifying this point with lattice simulations.}  This implies under--density of the DM quanta at a given temperature and thus the presence of negative chemical potential.

The thermalization of dark matter is controlled by the self--coupling $\lambda$. For  small $\lambda \ll 1$, e.g. $10^{-6}$ for typical cases, dark matter reaches kinetic equilibrium at some stage, however the number changing processes are suppressed by a further factor of $\lambda^2$ and chemical equilibrium never sets in.  

Note that integrating out the inflaton leads to a tiny Higgs--DM coupling of order $ (\lambda_{\phi S} /8\pi^2) \sigma^2 /m_{\phi}^2 $.  At sufficiently small $\sigma$, it is irrelevant to both thermalization and freeze--in production of dark matter \cite{Hall:2009bx}, making the effect emphasised   in Ref.\;\cite{Adshead:2016xxj} negligible. 

 This example illustrates that dark matter and observed matter can be produced by very different mechanisms, in which case one expects different temperatures in the two sectors. Furthermore, the dark sector can be endowed with effective chemical potential,  as long as the number changing interactions are inefficient.

\section{Boltzmann equation and reaction rates}

\begin{figure}[h]
\begin{center}
\includegraphics[scale=1]{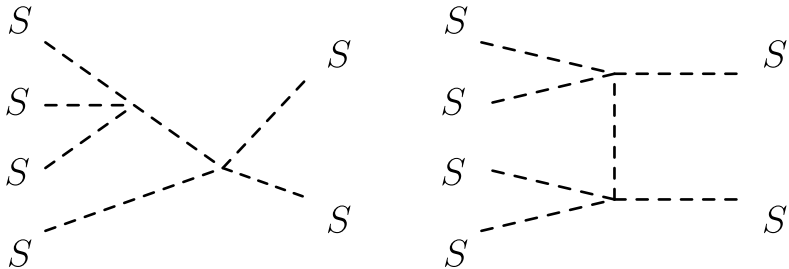}
\end{center}
\caption{Lowest order number changing processes.
\label{fig:diagram}}
\end{figure}

The  particle density  $n(t)$ evolution is  described by the Boltzmann equation. 
In addition to the Universe expansion, $n(t)$ is  affected by the particle number changing processes 
such as $SS \leftrightarrow SSSS$  (Fig.~\ref{fig:diagram}) and those of higher order.
Keeping the lowest order terms,   the Boltzmann equation in the FRW background reads 
(see e.g. \cite{Kolb:1990vq,Bernal:2015bla})
\begin{equation}
{dn \over dt} +3Hn = 2 ~(\Gamma_{2\rightarrow 4} -\Gamma_{4\rightarrow 2} )\;,
\label{boltzmann}
\end{equation}
%%%%%%%%%%%%%%%%%%%%%%%%%%%%%
%%%%%%%%%%%%%%%%%%%%%%%%%%%%%%
where  $H=\dot{a} /a$,  the factor of 2 comes from the particle number change in the scattering process and the reaction rates
per  unit volume  are 
\begin{equation}
\Gamma_{a\rightarrow b} = \int \left( \prod_{i\in a} {d^3 {\bf p}_i \over (2 \pi)^3 2E_{i}} f(p_i)\right)~
\left( \prod_{j\in b} {d^3 {\bf p}_j \over (2 \pi)^3 2E_{j}} (1+f(p_j))\right)
\vert {\cal M}_{a\rightarrow b} \vert^2 ~ (2\pi)^4 \delta^4(p_a-p_b) . 
\label{Gamma}
\end{equation}
Here ${\cal M}_{a\rightarrow b}$ is the  QFT transition amplitude,  in which we also absorb the {\it  initial and final} state symmetry factors;  
$f(p)$ is the momentum distribution function. It can deviate from the corresponding thermal distribution,
yet
as long as the system enjoys {\it kinetic} equilibrium through efficient $2\rightarrow 2$ scattering, $f(p)$ takes the form  
 \cite{Bernstein:1985th},\cite{Kolb:1990vq}
\begin{equation}
f(p)= {1 \over \exp^{E-\mu\over T} -1 } \;,
\end{equation}
where $\mu$ is the {\it effective} chemical potential. This can be understood from the fact that such a Bose--Einstein distribution maximizes the entropy with the approximately constant particle number, while the number changing interactions 
are relatively slow.

In thermal (chemical) equilibrium,  the number changing reaction rates are much greater than the Hubble rate $H$ and   $\Gamma_{2\rightarrow 4} $ equals exactly $\Gamma_{4\rightarrow 2}$. This is because at $\mu=0$
\begin{eqnarray}
&& f(k_1)~f(k_2)~ f(k_3)~f(k_4)~ \bigl(1+f(p_1)\bigr) ~\bigl(1+f(p_2)\bigr) = \nonumber\\
&& f(p_1)~f(p_2)~ \bigl(1+f(k_1)\bigr)~ \bigl(1+f(k_2)\bigr) ~\bigl(1+f(k_3)\bigr) ~\bigl(1+f(k_4)\bigr) 
\end{eqnarray} 
due to energy conservation in the reaction $p_1p_2 \leftrightarrow k_1k_2k_3k_4$, and $\vert {\cal M}_{a\rightarrow b} \vert =\vert {\cal M}_{b\rightarrow a} \vert$. 
As a result, the right hand side of the Boltzmann equation vanishes and the total number of particles is conserved,
\begin{equation}
{d \over d t} n a^3 =0 \;. 
\end{equation}

When the system departs from thermal equilibrium and a non--zero $\mu$ develops, the cancellation is no longer exact. 
As a result, the total particle number changes.
Eventually, due to the Universe expansion,  both $\Gamma_{2\rightarrow 4} $ and  $\Gamma_{4\rightarrow 2}$ decrease to the extent  that the right hand side of the Boltzmann equation becomes negligible again and the total particle number remains approximately constant. This happens roughly when
 \begin{equation}
3Hn \gtrsim 2 ~\Gamma_{2\rightarrow 4} \; , \; 2~\Gamma_{4\rightarrow 2} \;,
\end{equation}
which is called ``freeze--out''.  The freeze--out can take place in both relativistic and non--relativistic regimes, which we will consider separately in what follows.

In order to understand both regimes, we need to derive compact expressions for the rates which can be analyzed either analytically or numerically.

\subsection{Relativistic reaction rates with the Bose--Einstein distribution function }

Consider the reaction rate $\Gamma_{2\rightarrow 4} $. Following Gelmini and Gondolo  \cite{Gondolo:1990dk}, we find it
 convenient to express this rate in terms of the cross section $\sigma (p_1,p_2)$,
 \begin{equation}
\sigma (p_1,p_2)= {1\over 4 F(p_1,p_2)} \int \vert {\cal M}_{2\rightarrow 4} \vert^2 (2\pi)^4 \delta^4(p_1+p_2 -\sum_i k_i)
\prod_i {d^3 {\bf k}_i \over (2 \pi)^3 2E_{k_i}} (1+f(k_i)) \;,
\label{sigma-def}
\end{equation} 
where $F(p_1,p_2) = \sqrt{(p_1 \cdot p_2)^2 - m^4} $ and $1+f(k_i)$ are the final state Bose--Einstein
enhancement factors. In this expression, the momentum distribution function can be put in a Lorentz--covariant form as (see e.g.\;\cite{Das:1997gg})
 \begin{equation}
f(p)= {1 \over e^{u\cdot p \over T} -1 } \;,
\end{equation} 
where $u_\mu$ is the 4--velocity of our reference frame relative to the gas rest frame in which $u=(1,0,0,0)^T$.
Apart from the Bose--Einstein factors, the above cross section is manifestly Lorentz--invariant.

The reaction rate is then expressed as
\begin{eqnarray}
&&\Gamma_{2\rightarrow 4}  =  (2\pi)^{-6} \int d^3 {\bf p_1} d^3 {\bf p_2} ~f(p_1) f(p_2) ~\sigma (p_1,p_2) v_{\rm M\o l}  \;,
 \label{Gamma-sigma}
 \end{eqnarray}
where the M\o ller velocity is defined by
 \begin{equation}
v_{\rm M \o l}= {F(p_1,p_2) \over E_1 E_2}   \;.
\end{equation}
It is clear that the rate is proportional to the thermal average $\langle \sigma v_{\rm M\o l} \rangle$.

The cross section is easiest calculated in the center--of--mass frame.\footnote{The public code CalcHEP \cite{Belyaev:2012qa} computes the cross sections either in the center--of--mass or in the lab frame.   Thus, for  numerical analysis, it is necessary to convert our expressions into one of these frames. }
Hence, for each pair $p_1,p_2$ in the gas rest frame, we find the   center--of--mass frame and  compute the 
corresponding cross section. The 4-velocity factor is thus a function of the momenta, $u_\mu  (p_1,p_2)$.

Introduce 
\begin{equation}
p={p_1+p_2\over 2} ~,~k={p_1-p_2\over 2} ~,
\end{equation}
such that 
\begin{equation}
{d^3 {\bf p_1}\over 2E_1} {d^3 {\bf p_2}\over2 E_2} = 
2^4 d^4p ~d^4k ~\delta\Big((p+k)^2-m^2\Big) ~\delta\Big((p-k)^2-m^2\Big)
\;.
\end{equation}
The center--of--mass frame is defined by the requirement that $p$ has zero spacial components. Let us parametrize the timelike $p$ as
 \begin{eqnarray}
&& p^0= E \cosh \eta, \nonumber\\ 
&& p^1=E \sinh \eta \sin \theta \sin \phi ,\nonumber\\
&& p^2=E \sinh \eta \sin \theta \cos \phi ,\nonumber\\
&& p^3=E \sinh \eta \cos \theta  ,\nonumber
\end{eqnarray}
where $E$ is the particle energy in the center--of--mass frame, $\eta$ is the rapidity and $\theta,\phi$ are
the angular variables (see Appendix A). Then
\begin{equation}
d^4p= \sinh^2 \eta  E^3dE ~d\eta ~d\Omega_p \;,
\end{equation}
where $\Omega_p$ is the solid angle in $p$-space. Due to the $\delta$--functions, the $k$--integral reduces to angular integration over the solid angle $\Omega_k$ in $k$--space. We thus have, for any $G(p_1,p_2)$, 
\begin{equation}
\int {d^3 {\bf p_1}\over 2E_1} {d^3 {\bf p_2}\over2 E_2} ~G(p_1,p_2)=
2 \int_m^\infty  dE    ~\sqrt{E^2-{m^2 }}  ~E^2 \int_0^\infty  d\eta  ~\sinh^2 \eta  ~ \int d\Omega_p ~d\Omega_k ~G(p_1,p_2) \;,
\end{equation}
where in the integrand one must set $k_0=0, \vert {\bf k}\vert=\sqrt{E^2-m^2}$ in $k$-dependent quantities. 
Note that $E$ is half the center--of--mass energy.

The cross section is  calculated in the center--of--mass frame. Thus, we need to transform the final state momenta $k_i$ to that frame, $k_i \rightarrow \Lambda k_i $, where $\Lambda$ is the corresponding Lorentz transformation given in Appendix A.  Due to Lorentz--covariance,  $\sigma(p_1,p_2)$
retains the same form except for the Bose--Einstein enhancement factors which now become
\begin{equation}
1+f(k_i)= 1+ {    1  \over e^{ (k_i^0 \cosh \eta + k_i^3 \sinh \eta- \mu)/T}  -1   }
\end{equation}
since $(\Lambda^{-1}u) \cdot k_i = k_i^0 \cosh \eta + k_i^3 \sinh \eta$ (see Appendix A).
As a result, the center--of--mass cross section  depends on $\eta$ as well,
 $\sigma_{\rm CM}(E,\eta)$. The angular integrations can be performed explicitly with the result
\begin{eqnarray}
&& \Gamma_{2\rightarrow 4} =  (2\pi)^{-6} \int d^3 {\bf p_1} d^3 {\bf p_2} ~f(p_1) f(p_2) ~\sigma (p_1,p_2) v_{\rm M\o l} = \label{Gamma24} \\
&& {4 T \over \pi^4} \int_m^\infty dE ~E^3 \sqrt{E^2-m^2} \int_0^\infty d\eta {    \sinh \eta \over e^{ 2(E  \cosh\eta -\mu)/T }-1}~
\ln {  \sinh    {E\cosh\eta + \sqrt{E^2 -m^2} \sinh\eta -\mu \over 2T}    \over
\sinh   {E\cosh\eta - \sqrt{E^2 -m^2} \sinh\eta -\mu \over 2T}  } \nonumber \\
&& \times \sigma_{\rm CM}(E,\eta) \;. \nonumber
\end{eqnarray}
Here,  $\sigma_{\rm CM}(E,\eta)$ includes the Bose--Einstein factors for the final state and is non--zero for $E\geq 2m$. The cross section can be computed numerically with CalcHEP by absorbing $1+f(k_i)$ into a momentum--dependent vertex. Note that in our convention, the symmetry factors due to the identical particles in the initial and final states, namely, $1/(2!4!)$,
have been absorbed into $\sigma$.
In the small $T$ limit, the final state enhancement factors can be neglected and  one recovers the Gelmini--Gondolo result  
\begin{eqnarray} 
&&\Gamma_{2\rightarrow 4} \simeq  {2 T\over \pi^4}  \int_m^\infty dE ~\sigma(E)~ E^2(E^2-m^2) ~ K_1(2E/T) ,
\label{MBresult}
\end{eqnarray}
where $K_1(x)$ is the modified Bessel function and $\mu$ is set to zero.

The rate $\Gamma_{4\rightarrow 2}$ can be obtained from $\Gamma_{2\rightarrow 4}$ by noting that 
\begin{equation}
f(p)=(1+f(p)) \; e^{- {u\cdot p -\mu\over T}} \;,
\end{equation}  
 such that 
  \begin{eqnarray}
&& f(k_1)~f(k_2)~ f(k_3)~f(k_4)~ \bigl(1+f(p_1)\bigr) ~\bigl(1+f(p_2)\bigr) = \nonumber\\
&& f(p_1)~f(p_2)~ \bigl(1+f(k_1)\bigr)~ \bigl(1+f(k_2)\bigr) ~\bigl(1+f(k_3)\bigr) ~\bigl(1+f(k_4)\bigr) \; e^{2\mu/T} \;.
\end{eqnarray} 
Therefore
\begin{equation}
\Gamma_{4\rightarrow 2} =  \Gamma_{2\rightarrow 4} \; e^{2\mu/T} \;.
\label{Gamma42Gamma24}
\end{equation}

 While obtaining  $\sigma_{\rm CM}(E,\eta)$ for the $2 \rightarrow 4$ and $4 \rightarrow 2$ reactions in a closed form does not seem possible,
 the $2 \rightarrow 2$ reaction is simple enough such that   $\sigma_{\rm CM}^{2 \rightarrow 2}(E,\eta)$ can be computed explicitly.
 The corresponding reaction rate is needed to describe kinetic equilibrium. 
 We have 
 \begin{equation}
 \sigma_{\rm CM}^{2\rightarrow 2}(E,\eta)=   {1\over 4F(p_1,p_2)} \int d\Omega ~ { \vert {\bf k_1} \vert \over (2\pi)^2 8 E }~
\vert  {\cal M} \vert^2 ~\Big(1+ f(k_1)\Big)~ \Big(1+f(k_2)\Big) \;,
\end{equation}
where  $\vert  {\cal M} \vert^2=\lambda^2/(2!2!)$. As explained above, we include the symmetry factors for the final and initial states in the amplitude,
so the large $E$ limit of this cross section differs from the standard result (see e.g.\;\cite{Peskin:1995ev})  by $1/2!$.   
This is admissible since we are only interested in the thermal averages, which are convention--independent.
  The angular dependence comes entirely from $f(k_i)$. Computing the integral, we obtain 
 \begin{equation}
  \sigma^{\rm CM}_{2\rightarrow 2}(E,\eta)= {1\over 2!2!} \times {\lambda^2 T \over 64 \pi E^2 \sqrt{E^2-m^2}  \sinh\eta}
~ {1 \over 1-e^{ -{2E \over T} \cosh \eta}} ~
\ln {  \sinh    {E\cosh\eta + \sqrt{E^2 -m^2} \sinh\eta \over 2T}    \over
\sinh   {E\cosh\eta - \sqrt{E^2 -m^2} \sinh\eta \over 2T}  }~,
\end{equation}
where we have set $\mu=0$. A non--zero $\mu$ is trivially included via the replacement $E\cosh\eta \rightarrow E \cosh\eta -\mu$.
Plugging this result into an analog of (\ref{Gamma-sigma}), we obtain $\Gamma_{2 \rightarrow 2}$. 

Finally, let us note that  at high temperatures it is important to include the thermal mass term, i.e.
\begin{equation}
m^2 \rightarrow m^2 + {\lambda\over 24 } T^2 \;.
\label{thermal-mass}
\end{equation}
 This regularizes the behaviour of the rates at $T\gg m$ and cures the infrared divergence as $m \rightarrow 0$.

\subsection{Conditions for thermal and kinetic equilibria}

Thermal or kinetic equilibrium is maintained if the relevant reaction rate is larger than the Hubble rate. It is convenient to express this condition in terms of the quantities appearing in the Boltzmann equation. For full thermal equilibrium, we require
\begin{equation}
3nH \lesssim 2 \Gamma_{2\rightarrow 4} \;, 
\label{ther-eq}
\end{equation}
which implies that the number changing interactions are efficient and the Bose--Einstein distribution is realized. On the other hand, kinetic equilibrium is maintained as long as
  \begin{equation}
3nH \lesssim  \Gamma_{2\rightarrow 2} \;,
\end{equation}
such that the scattering rate is high enough to define a temperature. This is a weaker condition since normally $\Gamma_{2\rightarrow 2} \gg \Gamma_{2\rightarrow 4}$. The system is still described by the Bose--Einstein distribution which maximizes  entropy,  however 
a non--zero chemical potential is necessary to account for approximate particle number conservation.

 \begin{figure}[h]
\begin{center}
\includegraphics[scale=1]{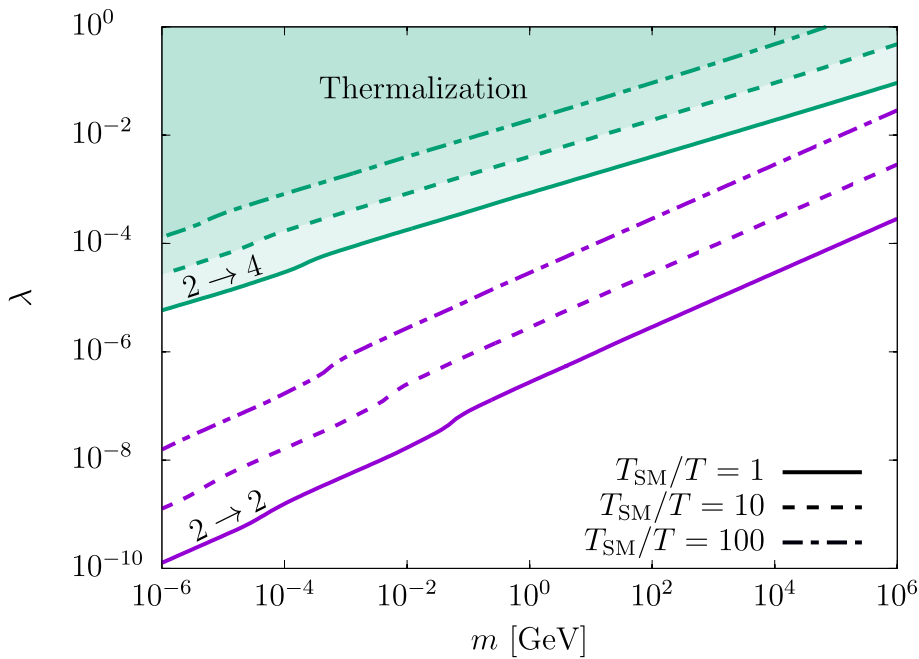}
\end{center}
\caption{Parameter space consistent with  thermal or kinetic equilibrium for different $\xi=T_{\rm SM}/T$. Couplings above the green (purple) lines are necessary for thermal (kinetic) equilibrium neglecting the chemical potential.
\label{fig:thermalization}}
\end{figure}

Throughout this paper we assume that before and around freeze--out, the energy density of the Universe is dominated by the SM fields. This can be either due
to higher temperature $ T_{\rm SM}$  in the observed sector or due to  a larger number of  SM degrees of freedom $g_*$. In this case,
\begin{equation}
  a= {{\rm const} \over T_{\rm SM}} ~~,~~H= \sqrt{\pi^2 g_* \over 90}~ {T_{\rm SM}^2\over M_{\rm Pl}} \;.
 \end{equation}
In the relativistic regime, $T \gg m$, the temperature ratio
\begin{equation}
\xi \equiv {  T_{\rm SM} \over T}
\end{equation}
remains constant. 
We find that the ratio $\Gamma_{2\rightarrow 4} /(nH)$ is maximized around $T \sim 5m/\sqrt{\lambda}$, i.e. when the thermal mass becomes comparable 
to the bare mass (see Sec.\;\ref{sec-UR-FO}). Imposing  condition (\ref{ther-eq}) at this temperature, we  obtain
a lower bound on $\lambda$ at each $m$. The  allowed parameter space is shown in Fig.~\ref{fig:thermalization}. The values of $\lambda$ above the green lines are $consistent$ with thermal equilibrium for a given $\xi$.\footnote{The constraint provides the necessary condition for thermal equilibrium, while the thermalization process depends on further details such as the initial distribution function.} (Naively,  since $n \propto T^3$ and $\Gamma_{2\rightarrow 4} \propto   T^4$, 
 one expects the constraint    (\ref{ther-eq})  at high temperature to be  of the type
$ m \ll T<  {\rm const} \times \lambda^4 \xi^{-2} M_{\rm Pl} $; 
 while  this shows the right trend, i.e. the minimal $\lambda$ increases with $m$,    in reality the $\lambda$--dependence is more complicated due to the thermal mass contribution.)

Regarding kinetic equilibrium, the ratio $\Gamma_{2\rightarrow 2} /(nH)$ is maximized at $T\sim {\cal O}(m)$. Using our result for $\Gamma_{2\rightarrow 2}$,
 we find the allowed parameter space  in Fig.~\ref{fig:thermalization}.\footnote{The reaction rate is affected by the chemical potential. However, if $|\mu|$ is small compared to the temperature, the effect is not very significant.}  The couplings above the purple lines are necessary for maintaining kinetic equilibrium in the relativistic regime. Since the rate involves a lower power of $\lambda$, the $m$ and $\xi$ dependences are  stronger than those in the case of full thermal equilibrium.

%Here we assume that thermal  equilibrium is reached naturally in the relativistic regime. In the non--relativistic case, the bound on the coupling may differ %somewhat, but typically it is stronger than the relativistic constraint. Given the inflationary pre-history, it would be rather difficult to arrange for the dark sector to stay  %non-relativistic, so our assumption of relativistic thermalization appears natural. 

Here we require thermalization in the relativistic regime. In the non--relativistic case, the bound on the coupling is stronger (see  Appendix B), so 
 Fig.\;\ref{fig:thermalization} gives the necessary condition.

We see that thermal equilibrium requires  self--coupling of order $10^{-4}$--$10^{-3}$ at $m\sim 1$ GeV, while for kinetic equilibrium $\lambda$ can be as small as $10^{-8}$--$10^{-7}$. Thus, there is a large window in which only kinetic equilibrium is maintained and the DM number is approximately conserved.

\section{Freeze--out}

In what follows, we consider separately the non--relativistic and relativistic freeze--out regimes.
Here we assume that the system enjoys full thermal equilibrium such that $\mu=0$ initially.
In this case, the relic DM density is affected by the number changing interactions, which we
study in detail.

\subsection{Non--relativistic freeze--out}

In the non--relativistic regime, the expressions for the reaction rates simplify. 
The momentum distribution is given by the Maxwell--Boltzmann function $f(p)= e^{- (E-\mu)/T}$
and the final state enhancement factors can be neglected. 
In this case, the chemical potential dependence factorizes out
and according to (\ref{Gamma}) we have, 
\begin{equation}
\Gamma_{2\rightarrow 4}= 
e^{-2 \mu/T} \Gamma_{4\rightarrow 2}= e^{2 \mu/T} \Gamma_{4\rightarrow 2}(\mu=0)  \;.
\end{equation}
It is conventional to  $define$
\begin{equation}
\sigma_{4\rightarrow 2}v^3 \equiv  
{1\over 2 E_{k_1}    2 E_{k_2} 2 E_{k_3} 2 E_{k_4}     } \int {  d^3{\bf p_1} \over  (2\pi)^3 2 E_1 } {  d^3{\bf p_2} \over  (2\pi)^3 2 E_2 }
\vert {\cal M}_{4\rightarrow 2} \vert^2 ~ (2\pi)^4 \delta^4\left( \Sigma p_i- \Sigma k_j\right) \;,
\label{sigmav3}
\end{equation}
which is momentum independent in the non-relativistic limit $k_i \simeq (m,\vec{0})^T$. As usual, we absorb the symmetry factor $1/(2!4!)$ into 
$\vert {\cal M}_{4\rightarrow 2} \vert^2$.
We thus have 
\begin{equation}
\Gamma_{4\rightarrow 2}(\mu=0)= 
  (2\pi)^{-12}   \int   \prod_i \left( d^3{\bf k_i}~ e^{-E_{k_i}/T} \right) \sigma_{4\rightarrow 2}v^3=
  \langle    \sigma_{4\rightarrow 2}v^3  \rangle  n_{\rm eq}^4 \;,
   \end{equation}
where the equilibrium particle density  is $ n_{\rm eq} =(2\pi)^{-3} \int d^3 {\bf p} f(p)_{\mu=0}$   and  $\langle ... \rangle$ denotes a thermal average at $\mu=0$ over the momenta of the incoming particles. Expressing the chemical potential in terms of the particle densities as $e^{\mu/T} = n/n_{\rm eq}$,
we obtain the Boltzmann equation in the form
 \begin{equation}
{dn \over dt} +3Hn = 2 \langle  \sigma_{4\rightarrow 2}  v^3 \rangle    
( n^2 n_{\rm eq}^2 -n^4) \;.
\label{boltzmann1}
\end{equation}
An important feature of this equation  is that $\langle  \sigma_{4\rightarrow 2}  v^3 \rangle $ is temperature independent.

It is convenient to replace the time variable with the SM sector temperature and the number density with the total particle number, 
which stays approximately constant when the number changing interactions become inefficient.
Let us define 
\begin{equation}
  x={ m \over T_{\rm SM} }~~,~~Y={n\over s_{\rm SM}} ~,
 \end{equation}
where $s_{\rm SM}$ is the entropy density dominated by the SM contribution,
 \begin{equation}
 s_{\rm SM}={2 \pi^2 \over 45}   {m^3\over x^3 }  g_{*s} \;, 
 \label{SSM}
 \end{equation}
 with $g_{*s}$ being the number of degrees of freedom contributing to the entropy. This number is a function of the temperature such that
 \begin{equation}
  {ds_{\rm SM} \over dx} = - {3s_{\rm SM}\over x}  \left(      1- {x\over 3 g_{*s}} {d g_{*s} \over dx}    \right) \;.
   \end{equation}
In terms of the new variables, the Boltzmann equation reads
\begin{equation}
 {dY\over dx} = - {  2\langle  \sigma_{4\rightarrow 2}  v^3 \rangle  s_{\rm SM}^3    \over   x \tilde H } \; \left(   Y^4-Y^2 Y^2_{\rm eq}   \right)\;,
  \end{equation}
where the modified Hubble rate is defined by
\begin{equation}
 \tilde H \equiv H \left(     1- {x\over 3 g_{*s}} {d g_{*s} \over dx}      \right)^{-1} \;,
 \label{H-tilde}
 \end{equation}
and $H= \sqrt{\pi^2 g_*/90} \; m^2/(x^2 M_{\rm Pl})   $. Observe that the  $x$--dependent prefactor on the right hand side of the Boltzmann equation
falls off sharply with $x$, namely as $x^{-8}$. This implies particle number conservation at late times. 

Our next task is to derive $Y_{\rm eq} (x)  $. Indeed, there are two unknowns in our system: $T(t)$ and $\mu(t)$ which should be determined by 2 equations.
The second constraint comes from entropy conservation in the dark sector, $s_{\rm } a^3 =const$, or
\begin{equation}
{s_{\rm }  \over s_{\rm SM}  }  \equiv c \;,
\end{equation}
which is constant in time.
The entropy density in the non--relativistic limit  is given by
\begin{equation}
  s_{\rm }= {m-\mu+T \over T} ~n ~~~,~~~ n=\left(  {mT \over 2\pi}  \right)^{3/2} e^{- {m-\mu\over T}} \;,
  \label{s-n}
   \end{equation}
where  in $ s_{\rm }$ we also include the subleading term proportional to $T$. 
This allows us to express $T$ as
\begin{equation}
 T=  {2\pi  T_{\rm SM}^2   \over m} \left(  {2\pi^2 \over 45} g_{*s}  Y \right)^{2/3} \exp\left[{{2\over 3} \left(  {c\over Y}-1 \right)} \right] \;.
 \label{T-Tsm}
 \end{equation}
Since $Y_{\rm eq} =Y e^{-\mu/T}  $, we obtain
\begin{equation}
 Y_{\rm eq}=Y \exp \left[  - {x^2 \over 2 \pi}   \left(  {2\pi^2 \over 45} g_{*s}  Y \right)^{-2/3}   \exp\left[   {2\over 3} \left(  1-{c\over Y} \right)  \right]    + {c\over Y} -1 \right] \;.
 \end{equation}
This is to be inserted in the Boltzmann equation, while $c$ is determined by the boundary condition at $\mu=0$:
\begin{equation}
  c=   \left(   {m\over T_0}  +1\right) \;Y_0\;,
   \end{equation}
and  $Y_0$ is fixed  by the initial dark and observed sector  temperatures, $T_0$ and $T_{{\rm  SM}0}$.

\begin{figure}[h!]
\begin{center}
\includegraphics[scale=0.63]{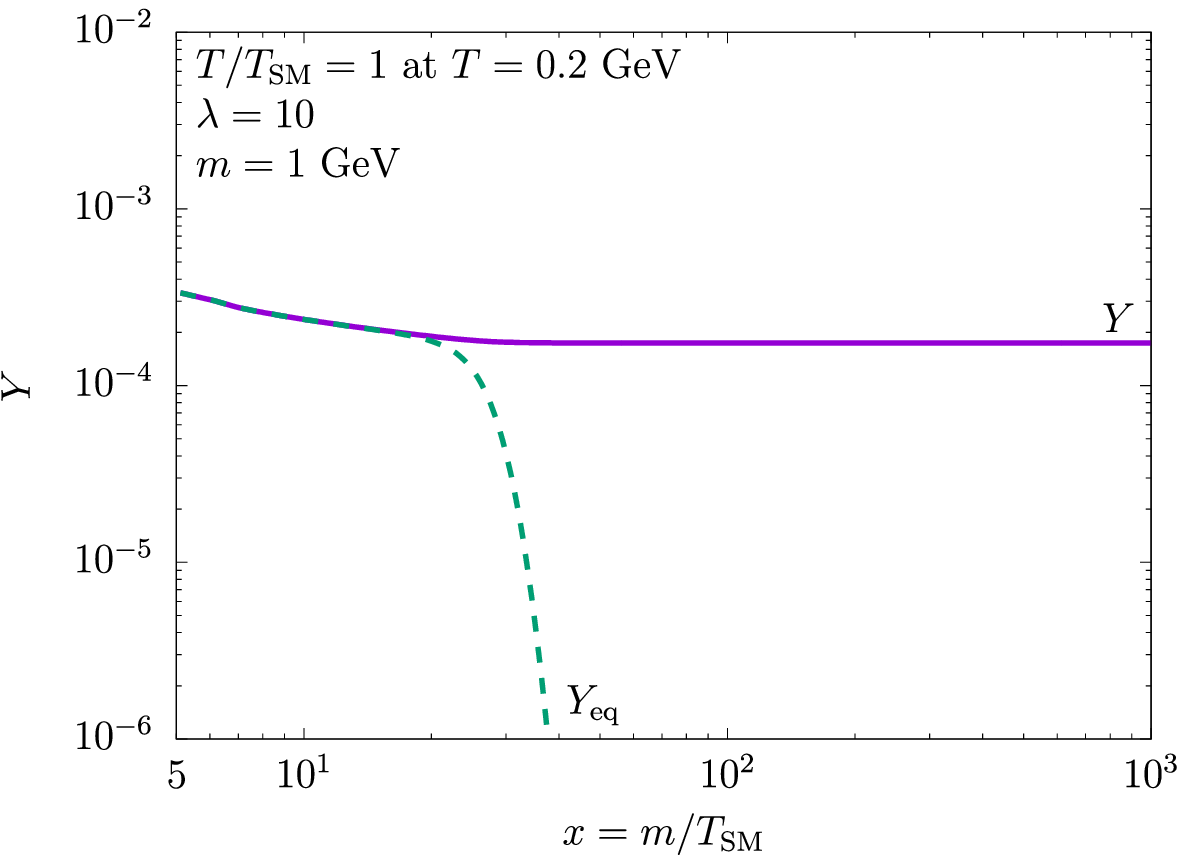}
\includegraphics[scale=0.63]{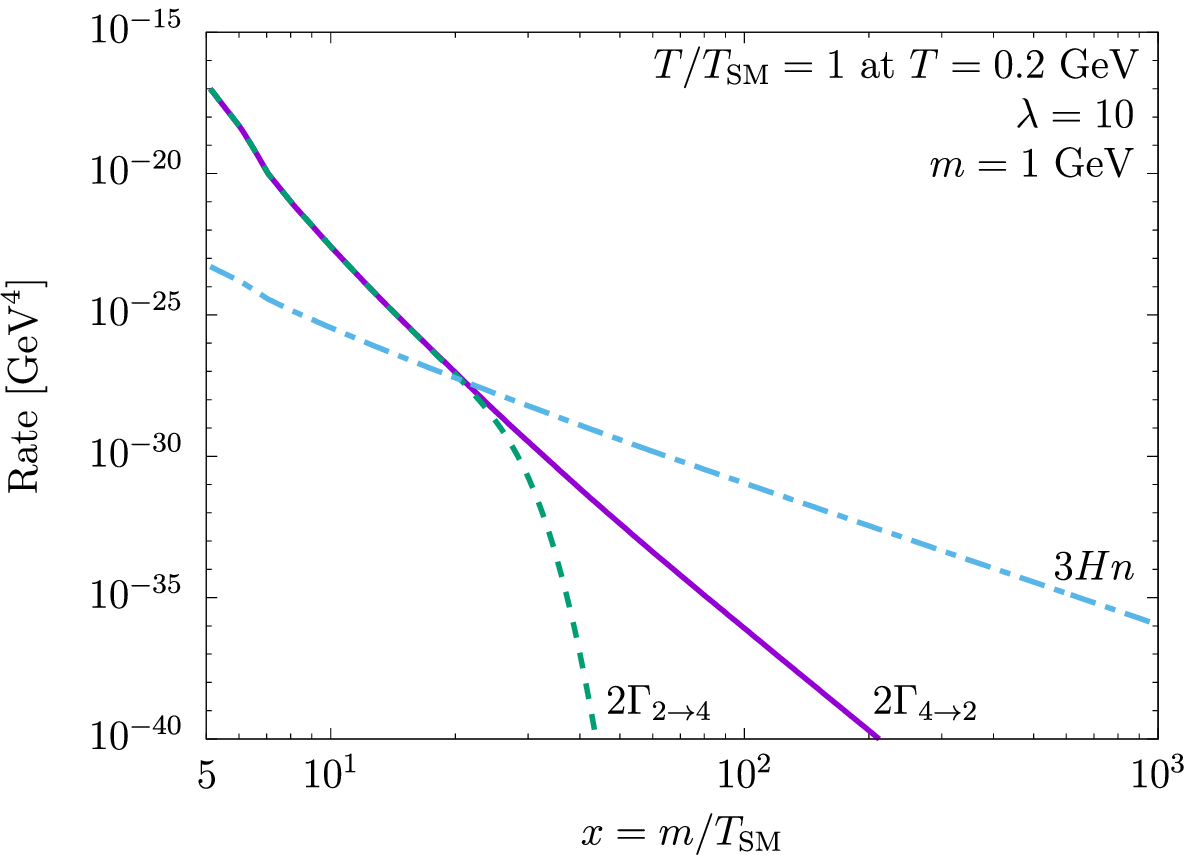}
\includegraphics[scale=0.63]{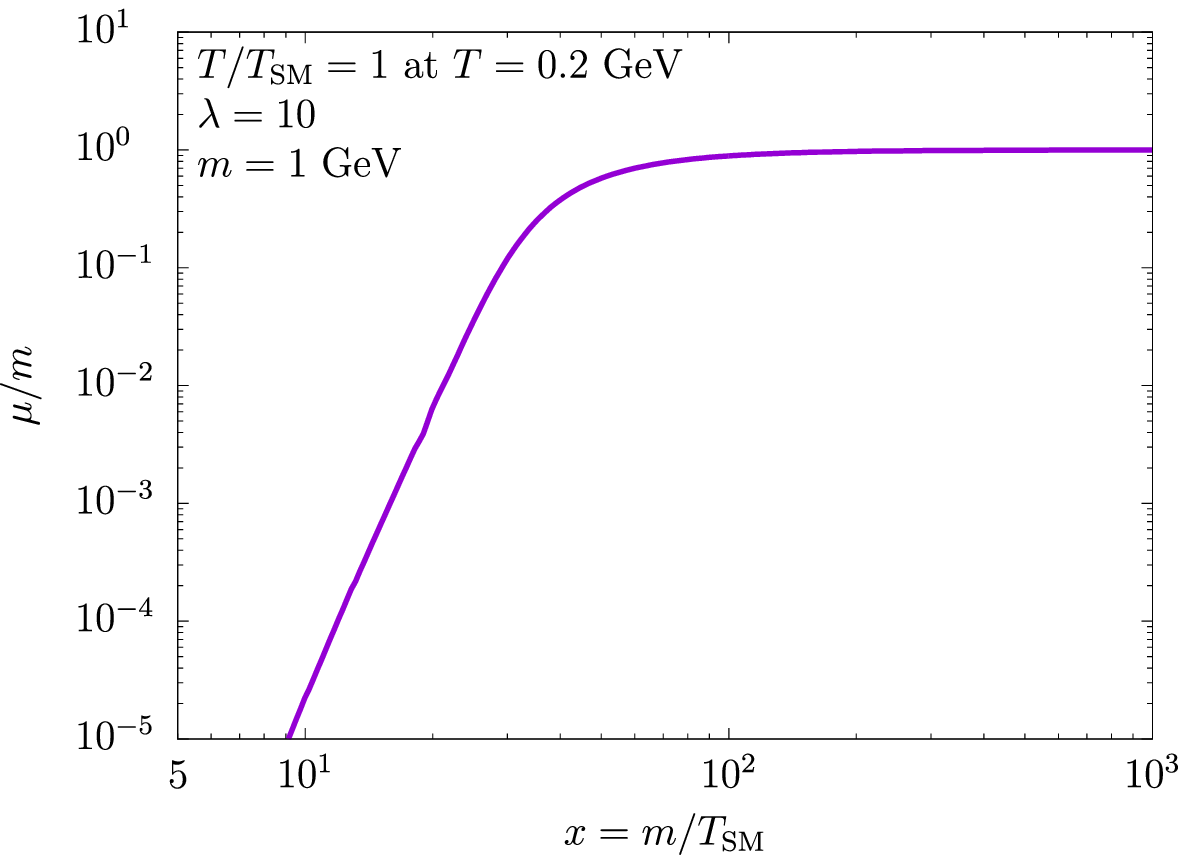}
\includegraphics[scale=0.63]{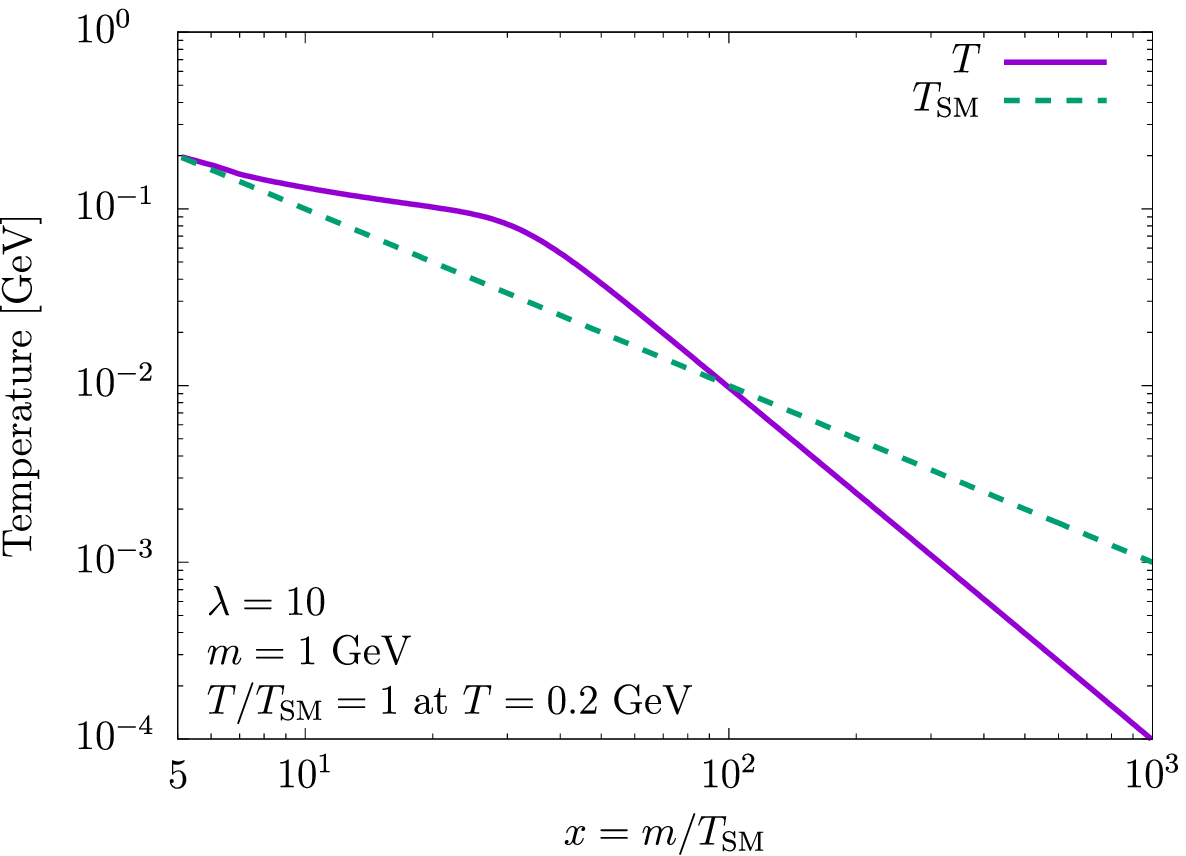}
\end{center}
\caption{  Thermodynamic quantities as a function of $x$ for non--relativistic freeze--out.
\label{fig:NR}}
\end{figure}

Now the Boltzmann equation can be solved numerically.
We assume that at the initial point  defined by $T_0$ and $T_{{\rm  SM}0}$, the system enjoys thermal equilibrium ($\mu=0$). Then, $Y(x)$ is found 
by solving the Boltzmann equation with this boundary condition.
In the non--relativistic limit, we find (see Appendix C)
 \begin{equation}
{\langle  \sigma_{4\rightarrow 2}  v^3 \rangle =   { \sqrt{3} \lambda^4 \over 2!4! \;256 \pi m^8 } \;, } 
\label{NRsigma}
\end{equation}
where we have factored out the $1/(2!4!)$ symmetry coefficient associated with the initial and final  state phase space. The resulting solution for a representative set of input 
parameters is shown in Fig.~\ref{fig:NR}.
 
%%%%%%%%%%%%%%%%%%%%%%%%%%%%%%%%%%
%%%%%%%%%%%%%%%%%%%%%%%%%%%%%%%%%%

The upper left panel of Fig.~\ref{fig:NR} shows that    $Y(x)$ follows $Y_{\rm eq}$ closely up until
$x\sim 25$, at which point it freezes--out and the number density remains approximately constant. As we see
from the right upper panel, the freeze--out is well described by
\begin{equation}
  3nH \simeq 2 \Gamma_{2\rightarrow 4} \simeq 2 \Gamma_{4\rightarrow 2} \;. 
 \end{equation}
After that, $\Gamma_{4\rightarrow 2} $ becomes suppressed compared to $3nH$, whereas $\Gamma_{2\rightarrow 4} $
turns negligible even faster. 

The effective chemical potential becomes appreciable, of order $T$,    around the freeze--out point after which it can be approximated by a linear function of $T$  asymptotically approaching $m$,  $m-\mu \propto T$. This follows from the entropy and particle number conservation in the dark sector (see Eq.~\ref{s-n}). 
In this regime, $T \propto T_{\rm SM}^2$ as required by Eq.~\ref{T-Tsm}. Before the freeze--out, the $T$ dependence on $T_{\rm SM}$ is only logarithmic.
Non--relativistic behaviour of DM leads to the heating of the dark sector \cite{Carlson:1992fn}, which can be viewed as a result of appreciable  $\Gamma_{4\rightarrow 2} $ at that stage.

Note that  $Y_{\rm eq}(x)$ is not a solution to the Boltzmann equation since  $Y_{\rm eq}^\prime(x)$
does not vanish. The true solution  $Y(x)$ is close to the ``equilibrium'' value until freeze--out. This implies that the right hand side of the    Boltzmann equation  is necessarily non--zero which entails 
reduction of the total DM number $na^3$ during this period. We find that the reaction rate difference
 $ 2(\Gamma_{4\rightarrow 2} - \Gamma_{2\rightarrow 4})$ is indeed not far from $3nH $ such that the
 number reduction is tangible.  
 
After the freeze--out, the particle number is approximately constant, typically within 10\%. Hence one can approximate
\begin{equation}
   Y(\infty) \simeq Y(x_{f}) \;,
 \end{equation}
 where $x_f$ is the freeze--out point.
 This is a slightly different condition compared to what is often used in the literature. Here, we do not neglect the  $Y_{\rm eq}(x)$ term which makes the number reduction less efficient, especially in the vicinity
 of the freeze--out point.

\subsection{Relativistic freeze--out} 

\begin{figure}[h!]
\begin{center}
\includegraphics[scale=0.63]{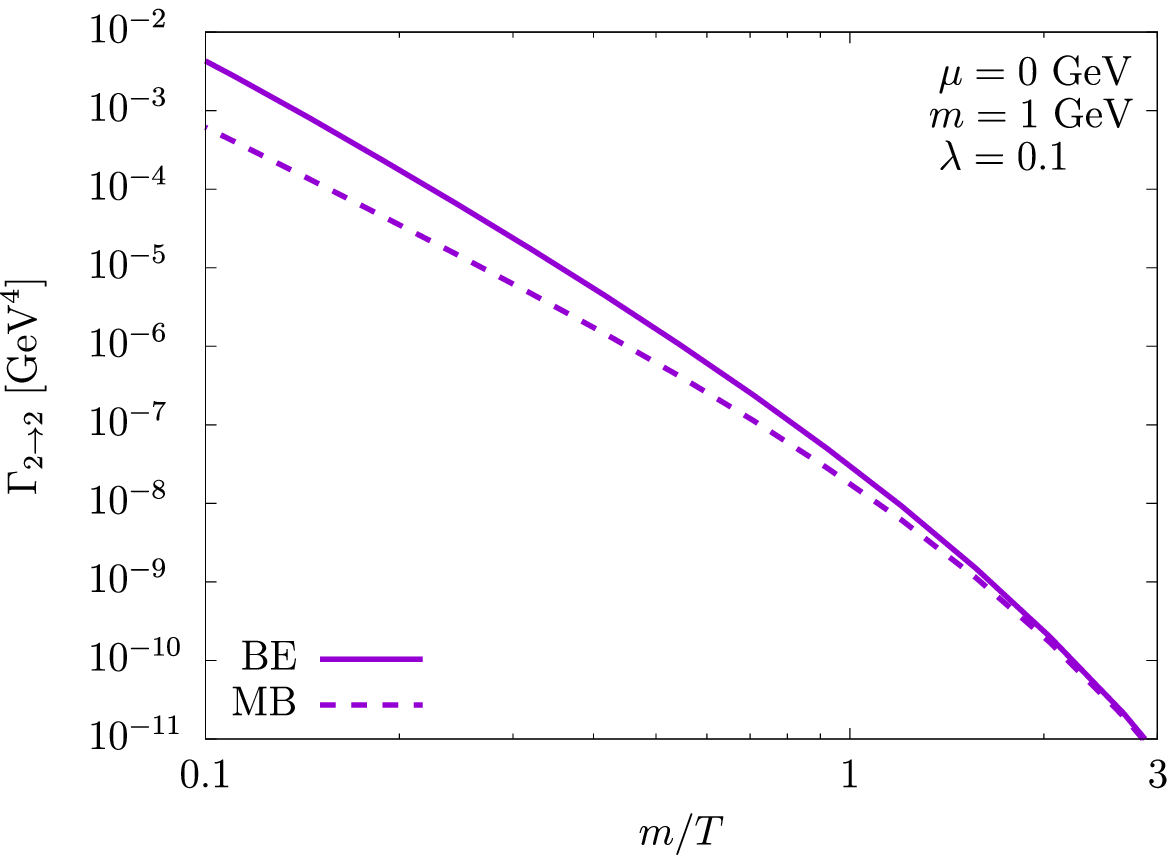}
\includegraphics[scale=0.63]{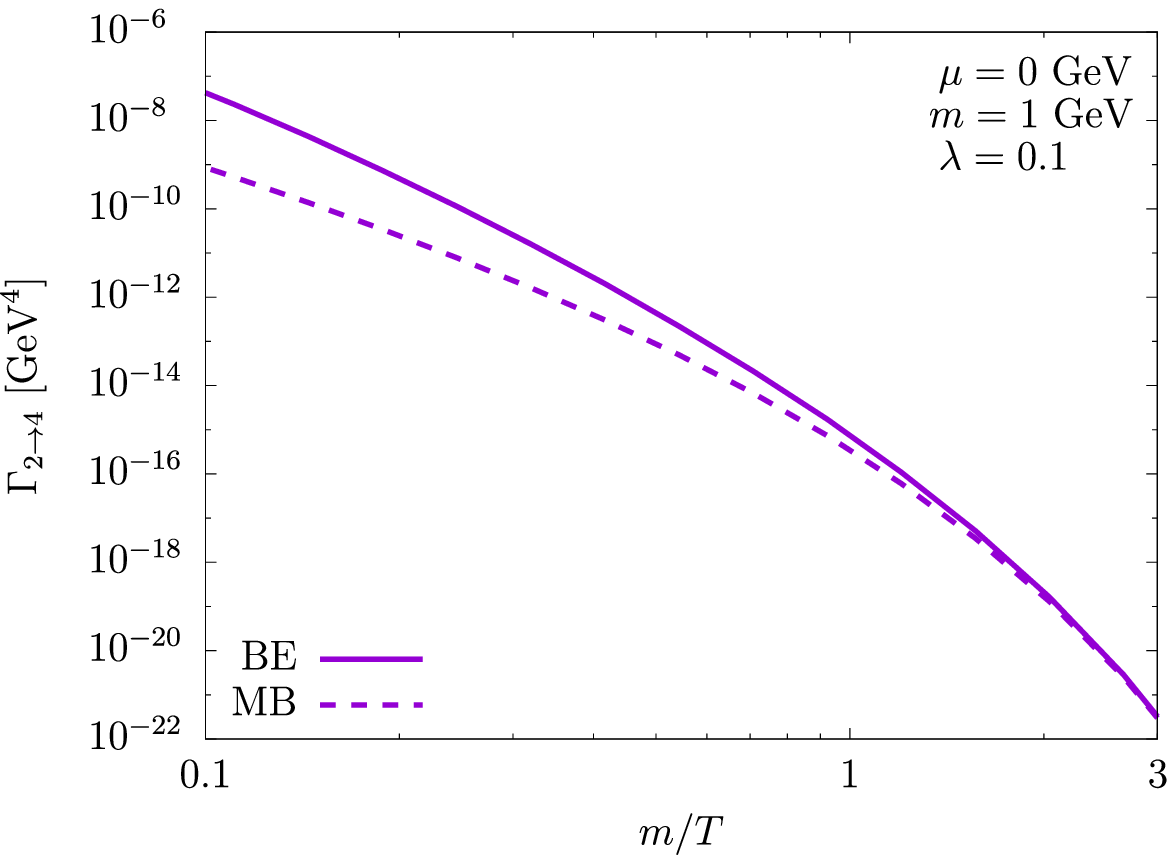}
\end{center}
\caption{   Reaction rates with Bose--Einstein statistics versus those with Maxwell--Boltzmann statistics.
\label{fig:MBvsBE}}
\end{figure}

For $m/T \lesssim 1$, the relativistic effects are important and one should use the Bose--Einstein distribution function. As Fig.\;\ref{fig:MBvsBE} shows,
the resulting reaction rates differ from their Maxwell--Boltzmann analogs by $10\%$ to $100\%$ at $m/T \sim 1$, while at 
$m/T \sim 0.1$ this difference can reach two orders of magnitude. The Bose--Einstein rates are greater due to the 
 enhancement factors for the low energy states. 
Therefore, the effect  is sensitive to the value of the thermal mass and $decreases$  for larger couplings.

%%%%%%%%%%%%%%%%%%%%%%%%%%%%%%%%%%
%%%%%%%%%%%%%%%%%%%%%%%%%%%%%%%%%%

As in the non--relativistic case, the evolution of the system is determined by the chemical potential and the temperature.
The Boltzmann equation and entropy conservation fix $\mu$ and $T$ as functions of $T_{\rm SM}$. Defining
\begin{equation}
y \equiv {m \over T}
\end{equation}
and $x \equiv m/T_{\rm SM }$, we rewrite the Boltzmann equation as an equation for $\mu(y)$ and $x(y)$:\footnote{This can be derived by 
expressing the left hand side of the Boltzmann equation in terms of $dY/dx$ with  $Y=n/s_{\rm SM}$, and calculating this derivative explicitly in terms of $f(p)$.}
\begin{equation}
  {1\over m} \;{d\mu \over  d \log y} = {  I_1 \over  I_2} + {1\over y \tilde H  I_2} \;{d \log x \over d \log y  } \; \left[    2(\Gamma_{2\rightarrow 4}-\Gamma_{4\rightarrow 2} ) -3Hn   \right] \;,
 \label{Boltzmann-rel}
 \end{equation}
where
\begin{eqnarray}
&& I_1  =\int {d^3 {\bf p}\over (2\pi)^3} f(p) (1+f(p)) \left[   {E-\mu \over m} - {\lambda T^2 \over 24 mE}  \right] \;,   \label{I_1} \\
&&  I_2 =\int {d^3 {\bf p} \over (2\pi)^3} f(p) (1+f(p))   \;,
\end{eqnarray}
with $E(p) =   \sqrt{ m^2_{\rm eff} + {\bf p}^2} \equiv   \sqrt{ m^2 + \lambda T^2/24 + {\bf p}^2}$, 
$n=\int d^3 {\bf p} /(2\pi)^3 f(p)$
and $I_{1,2} , \Gamma_{2\rightarrow 4}$,  $\Gamma_{4\rightarrow 2} $,  $\tilde H, n$ are to be expressed in terms of $\mu,x$ and $y$.
The modified Hubble rate $\tilde H$ is defined by Eq.\;\ref{H-tilde}. Note that we have included the thermal mass correction which leads to an extra contribution in Eq.\;\ref{I_1}.

The second equation for $\mu(y)$ and $x(y)$ is provided by the entropy conservation condition
\begin{equation}
{s\over s_{\rm SM} } = {\rm const}
\end{equation}
with
\begin{equation}
s= {\rho+p -\mu n \over T} 
\end{equation}
and $s_{\rm SM}$ from Eq.\;\ref{SSM}.
Here the energy density $\rho$ and the pressure  $p$ are given by the standard Bose--Einstein formulas, and  are to be expressed in terms of $\mu$ and $y$.

%%%%%%%%%%%%%%%%%%%%%%%%%%%%%%%%%%
\begin{figure}[h!]
\begin{center}
\includegraphics[scale=0.63]{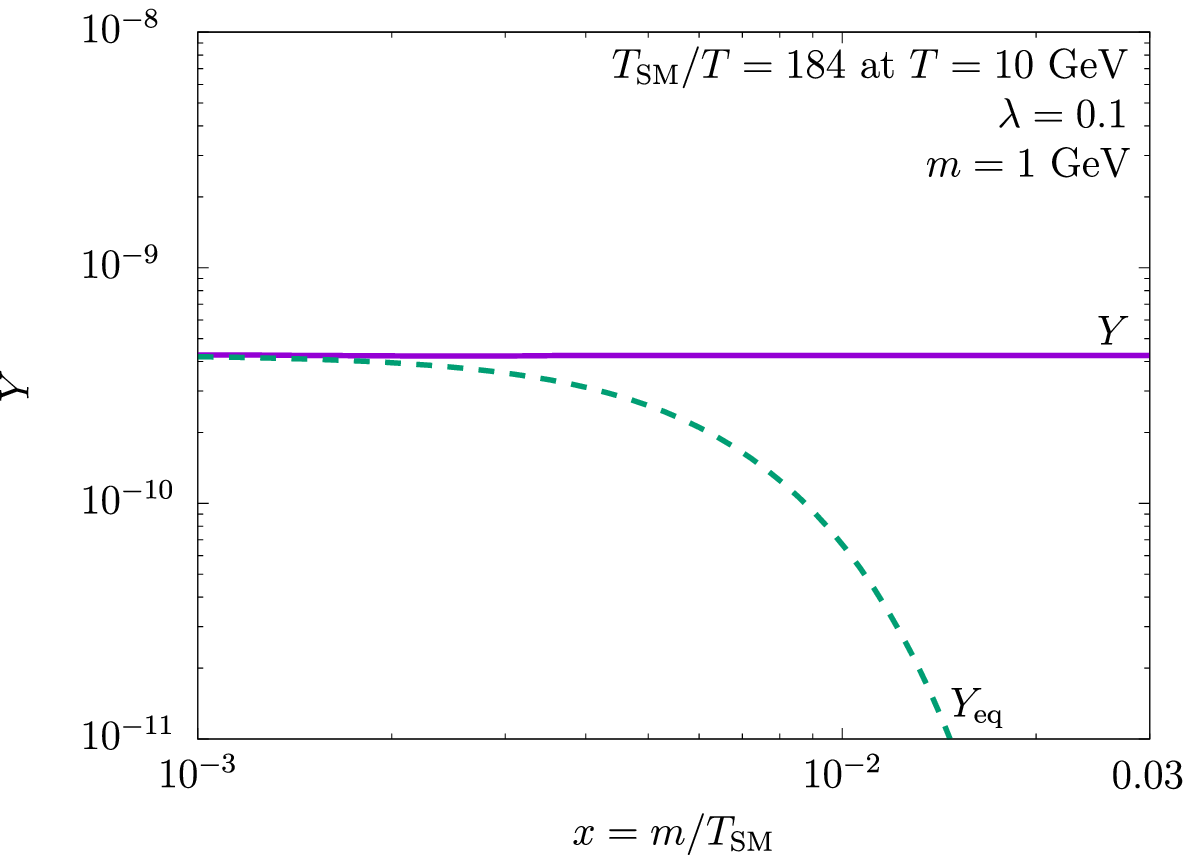}
\includegraphics[scale=0.63]{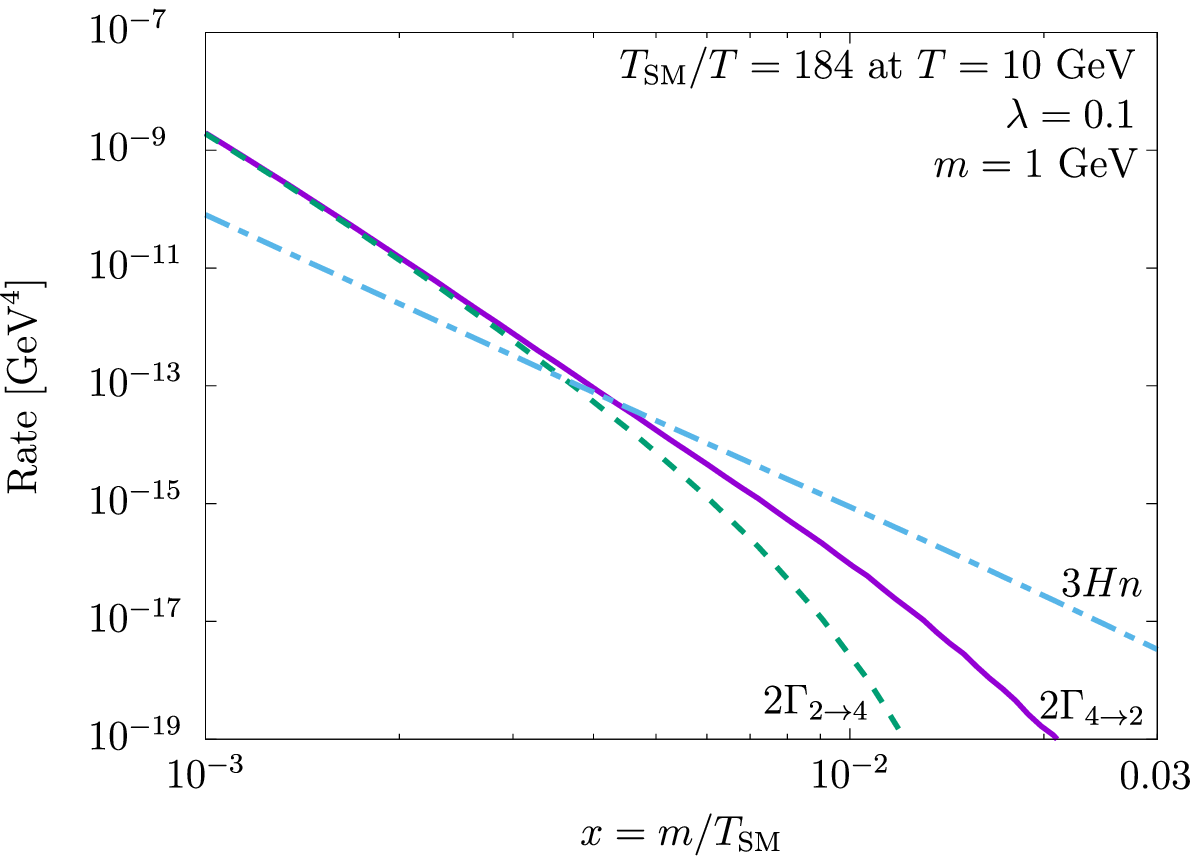}
\includegraphics[scale=0.63]{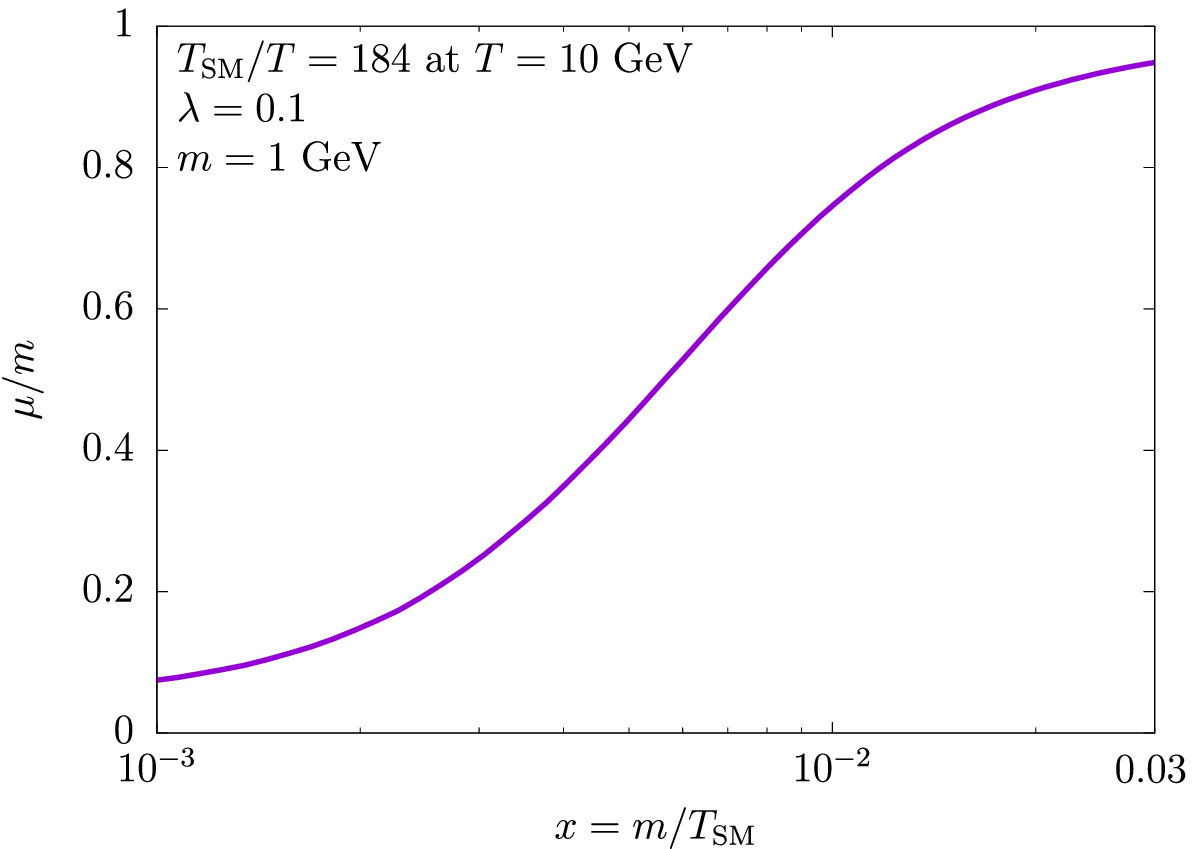}
\includegraphics[scale=0.63]{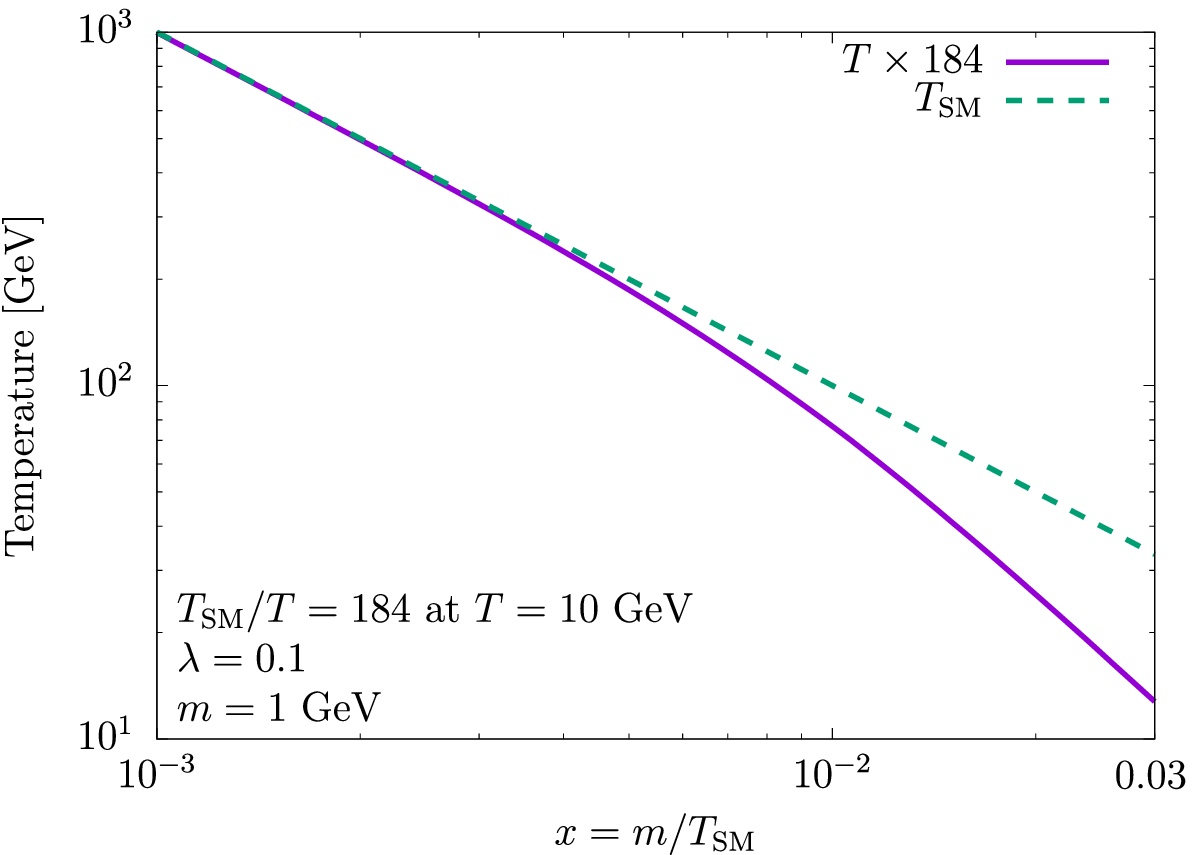}
\end{center}
\caption{  Thermodynamic quantities as a function of $x$ for relativistic freeze--out. Here $T_f \simeq 1.2$ GeV.
\label{fig:R}}
\end{figure}
%%%%%%%%%%%%%%%%%%%%%%%%%%%%%%%%%%

  %%%%%%%%%%%%%%%%%%%%%%%%%%%%%%%%%%
\begin{figure}[h!]
\begin{center}
 \includegraphics[scale=0.7]{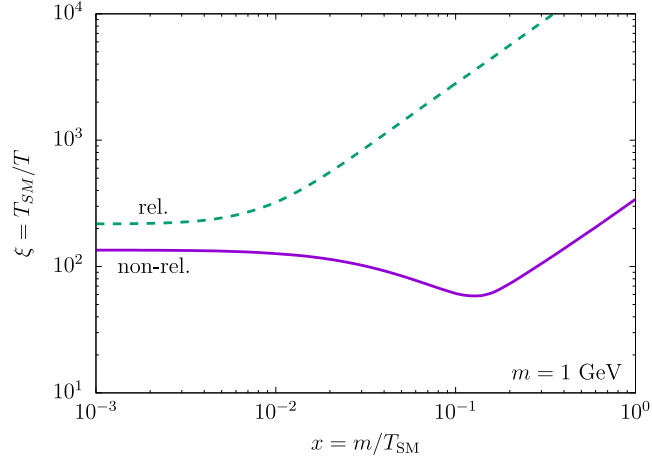}
\end{center}
\caption{   Evolution of $\xi=T_{\rm SM}/T$ for relativistic  and non--relativistic freeze--out. Dashed curve: $\lambda=0.1$ and $m/T_f =0.92$;
solid curve: $\lambda=10$ and $m/T_f =7.4$.
\label{fig:xi}}
\end{figure}
%%%%%%%%%%%%%%%%%%%%%%%%%%%%%%%%%%

The two coupled equations can be solved numerically.\footnote{We compute $\Gamma_{2\rightarrow 4}$ by  integrating numerically the CalcHEP output.} We present our results in Fig.\ref{fig:R}. The corresponding freeze--out temperature is $T_f=1.2$ GeV 
with $m=1$ GeV, which makes the freeze--out regime relativistic.

Compared to the non--relativistic case, we observe a few differences. First, the evolution of $Y$ and $\mu$ is slower. Second, there is no ``warming'' period in which
the dark temperature decreases much slower than $T_{\rm SM}$. After freeze--out, $T$ decreases faster than $T_{\rm SM}$ does. This is due to approximate conservation of the particle number  $n/T_{\rm SM}^3$: the increase in $\mu$ gets compensated by a decrease in $T$. Eventually, $T \propto T_{\rm SM}^2$ in the non--relativistic regime.  
For comparison, we present the evolution of $\xi$ for relativistic  and non--relativistic freeze--out in Fig.\;\ref{fig:xi}.

Again, we find that $Y(\infty)$ can be well approximated by $Y$ at freeze--out.

\subsection{Ultra--relativistic freeze--out}
\label{sec-UR-FO}

Freeze--out at $T \gg m, m/\sqrt{\lambda}$ is not possible in our model. In this regime, $T/T_{\rm SM}$  stays constant  by virtue of entropy conservation and
\begin{equation}
nH \propto T^5 ~~,~~ \Gamma_{2\rightarrow 4} \propto T^4 \;.
\end{equation}
Thus, if the thermal equilibrium condition $3nH < 2 \Gamma_{2\rightarrow 4} $ is satisfied at some point, it will continue to hold
as long as dark matter remains ultra--relativistic.\footnote{This is in contrast to the SM neutrino case, where the reaction rate involves a higher power
of $T$.} As a result, no freeze out is possible.

%%%%%%%%%%%%%%%%%%%%%%%%%%%%%%%%%%
\begin{figure}[h!]
\begin{center}
 \includegraphics[scale=0.6]{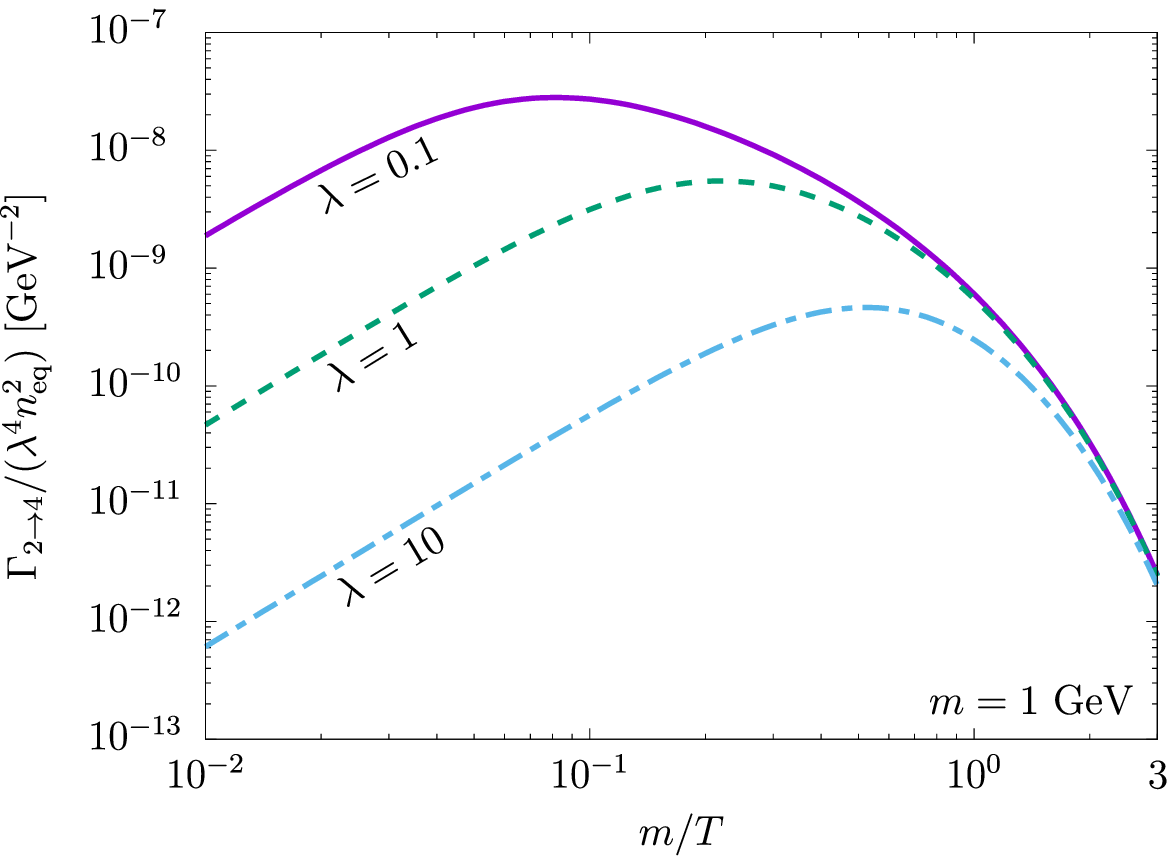} \;\;\;\;
  \includegraphics[scale=0.6]{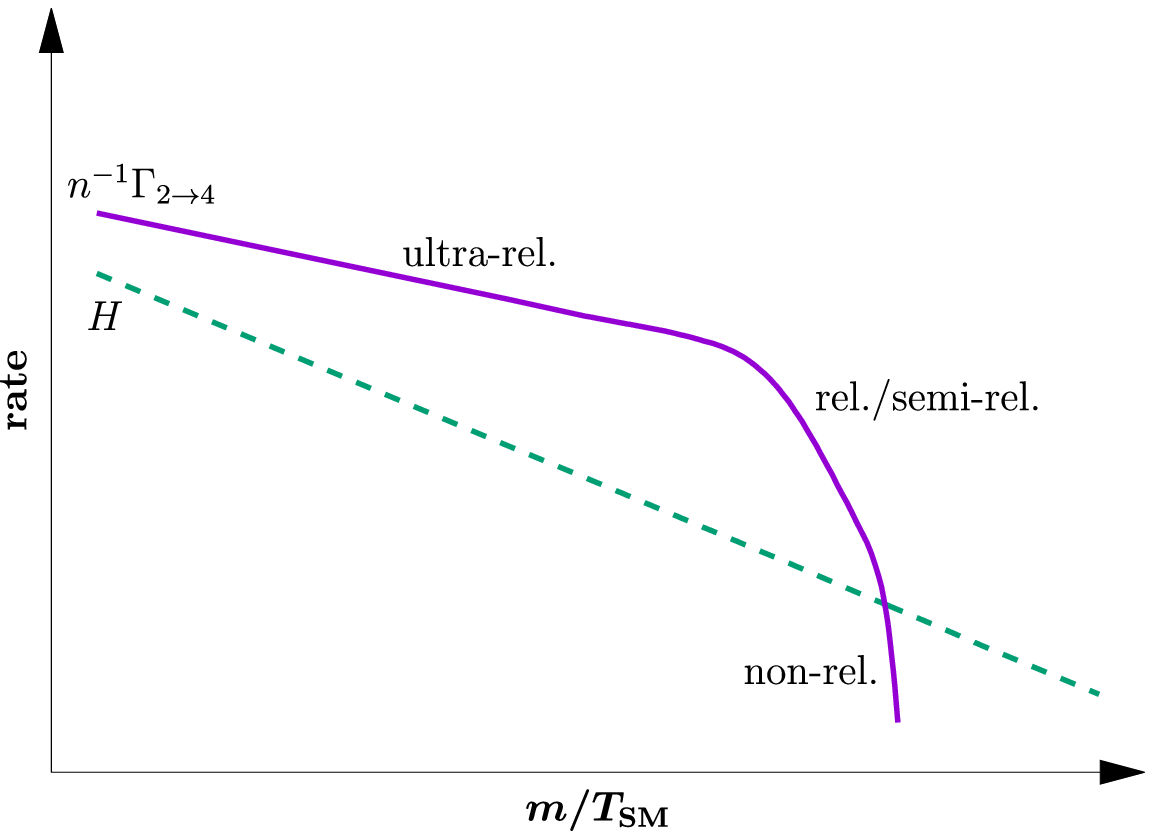}
\end{center}
\caption{  {\it Left:} Normalized $2\rightarrow 4$ rate dependence on the coupling. The thermal mass effect is clearly seen at $m/T \ll 1$. 
{\it Right:}  Log--scale evolution of $\Gamma_{2\rightarrow 4}/n$ and  $H$.
\label{fig:thermal-mass-effect}}
\end{figure}
%%%%%%%%%%%%%%%%%%%%%%%%%%%%%%%%%%

We note that the thermal mass effect is important in this regime. Due to the infrared singularity, the high temperature rates contain  the $T/m_{\rm eff}$ factors. For instance,
$\Gamma_{2\rightarrow 2} \propto T^4 \ln {T \over m_{\rm eff}} \;,$
where $m_{\rm eff}$ includes the thermal correction (\ref{thermal-mass}).
 At $T>5m/\sqrt{\lambda}$, the effective mass is dominated by the thermal term and the expected behaviour 
$\Gamma_{2\rightarrow 2} \propto T^4$ is reproduced.  (Here we neglect the usual log--running of the coupling constant which is  insignificant for  $\lambda$
in the range of interest.)
Similar considerations apply to $\Gamma_{2\rightarrow 4}$. The effect of the thermal mass is clearly seen in the left panel of Fig.\;\ref{fig:thermal-mass-effect}:
while in the non--relativistic regime the rate scales as $\lambda^4$, at higher temperatures this is no longer true. The $T^4$--behaviour is recovered when the thermal mass dominates.

Fig.\;\ref{fig:thermal-mass-effect}  (right) collects the different regimes in the $\Gamma_{2\rightarrow 4}$ behavior and presents an overall picture.
At high temperatures, $\Gamma_{2\rightarrow 4}/n \propto T$ evolves slower than $H\propto T_{\rm SM}^2$ does. This changes when the thermal mass becomes subdominant, which is marked as ``rel./semi-rel.'' in the plot. Finally, in the non--relativistic regime  $\Gamma_{2\rightarrow 4}/n \propto n^3$ is exponentially suppressed. The magnitude of the rate relative to $H$ is determined by the coupling constant. The plot makes it clear that freeze--out in the ultra--relativistic regime is impossible. Also,  if the dashed line is above the solid line, thermalization is never achieved. This is determined by $\Gamma_{2\rightarrow 4}$ in the 
relativistic/semi--relativistic regime where $T$ is not too far from $m$. The resulting lower bounds on $\lambda$ are presented in 
Fig.\;\ref{fig:thermalization}.

\section{Parameter space analysis}

In this section, we delineate our parameter space and determine regions consistent with the dark matter relic abundance constraint as well as other relevant bounds. We consider separately the full thermal and kinetic equilibrium cases.

 \subsection{Thermalized dark matter}

  %%%%%%%%%%%%%%%%%%%%%%%%%%%%%%%%%%
\begin{figure}[h!]
\begin{center}
 \includegraphics[scale=1.1]{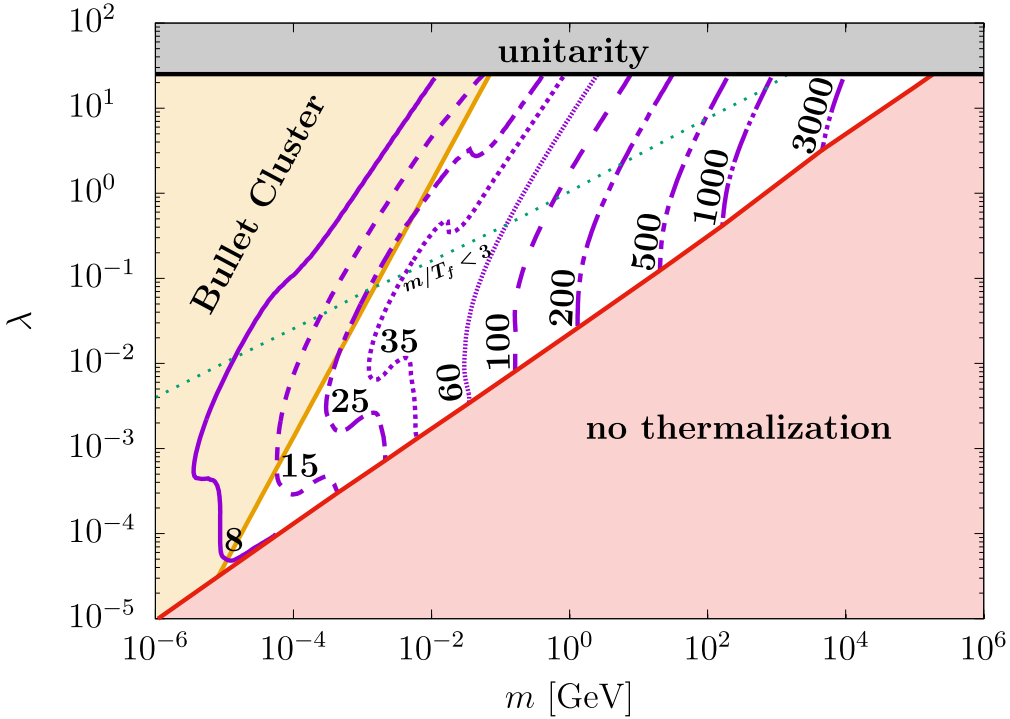}
\end{center}
\caption{  Parameter region (white) consistent with the correct DM relic density, Bullet Cluster, thermalization, and perturbative unitarity constraints. Points lying on each curve  reproduce the correct DM relic abundance for a specified $\xi(x_f)$ 
at freeze--out. In the red region, the relic density and freeze--out equations have no simultaneous solution.
\label{fig:par-space}}
\end{figure}
%%%%%%%%%%%%%%%%%%%%%%%%%%%%%%%%%%

  In this subsection, we assume that dark matter has been in thermal equilibrium and analyze the $(m,\lambda,\xi)$--parameter space 
    consistent with the observed DM relic density. This constraint can be put in the form
  \begin{equation}
  Y(\infty)=4.4 \times 10^{-10}~ \left(  {{\rm GeV} \over m } \right) \;.
  \end{equation}
  
  Let us discuss the main qualitative features of the model. 
  Consider first the non--relativistic freeze--out regime. As discussed earlier, we define the freeze--out point by
  \begin{equation}
  3nH = 2 \langle  \sigma_{4\rightarrow 2}  v^3 \rangle    n^4 \;.
  \end{equation}
  Solving this equation for $T_f$ and equating $Y(x_{f}) =Y (\infty)$, we find that the correct relic density is reproduced along the curves with the approximate scaling
  \begin{equation}
  \lambda \propto m \; \xi_f^{-7/4} \;,
  \end{equation}
  where $\xi_f$ is $T_{\rm SM}/T$ at the freeze--out point and we have neglected the {\it logarithmic} terms. For a fixed $\xi_f$, the mass--coupling relation is approximately linear and the freeze out temperature decreases with the coupling,
   \begin{equation}
   T_f \propto {m \over {\rm const } + \ln \lambda} \;,
     \end{equation}
     where the constant is positive.
  Thus, at sufficiently small $\lambda$, the non-relativistic approximation breaks down and a fully relativistic analysis is necessary.

  Our numerical results with  the relativistic reaction rates are presented in Fig.\;\ref{fig:par-space}. 
  We observe that in most allowed parameter space the freeze--out is relativistic, i.e. $m/T_f < 3$ as marked by the green dotted line.  In this region, the constant $\xi_f$ curves tend to approach a vertical  at $T \gtrsim m$ since $Y$ is determined mostly by $\xi$ and rather insensitive to $\lambda$.
  These curves exhibit characteristic kinks due to a change in the SM degrees of freedom at QCD phase transition ($T_{\rm SM} \sim 10^{-1}$ GeV) and electron decoupling. 
   It is interesting that $T_f$ does not vary monotonically along the constant $\xi_f$ curve: first, it decreases down to a minimum value and increases from there
    on. 

  For a given $\xi_f$, the correct relic density curve  cannot be continued indefinitely to smaller $\lambda$: at some point,  $3nH > 2\Gamma_{2\rightarrow 4}$ 
  for any $T$ (see Sec.\;\ref{sec-UR-FO}). Thus, the relic density and freeze--out equations have no simultaneous solution. 
  The excluded region is marked  ``no thermalization'' in Fig.\;\ref{fig:par-space}.  
  Although different points on the border of this region correspond to different $\xi$, 
  its shape is consistent with Fig.\;\ref{fig:thermalization}. 
  We note that here  the thermal mass effect is significant which makes   analytical calculations more challenging.  
  
  Another constraint is imposed by the bound on DM self--interaction from the Bullet Cluster, $\sigma/m <1 \;{\rm cm^2/g}$, with
 $\sigma = \lambda^2 /(128\pi m^2)$. It excludes light DM with significant self--coupling. Finally, perturbative unitarity is violated in the process $SS\rightarrow SS$
 if $\lambda \gtrsim 8 \pi $ \cite{Chen:2014ask}.
We note that the nucleosynthesis constraint on the effective number of neutrinos is insignificant here since the dark matter contribution to the energy density in the relativistic regime  is suppressed by $T^4/T^4_{\rm SM} \ll 1$.
     
     In the non--relativistic regime, we find qualitative agreement with the results of Ref.\;\cite{Bernal:2015xba} although there are numerical differences. 
     
   Fig.\;\ref{fig:par-space} shows that the correct relic density    can be obtained for a wide range of DM masses: from 10 keV to 100 TeV, as long as the dark temperature is significantly below $T_{\rm SM}$.  Thus, both the ``warm'' and the ``cold'' options are open.

\subsection{Dark matter in kinetic equilibrium and $\mu \not =0$ }

 For  small enough self--coupling (see Fig.\;\ref{fig:thermalization}), the system reaches only kinetic but not thermal equilibrium. The notion of temperature is still well defined in this case, but the Bose--Einstein distribution involves   an {\it effective} chemical potential from the start. The latter is determined by the initial number density.
Assuming $T\gg m$, we have
\begin{equation}
n= {T^3 \over \pi^3 } \;{\rm Li}_3(e^{\mu/T}) \;,
\label{n-mu}
\end{equation}
where ${\rm Li}_3(x)$ is the third degree polylogarithm, ${\rm Li}_3(x)= \sum_{n=1}^\infty 
x^n/n^3$. For a given initial $n$, $\mu$ is read off from (\ref{n-mu}).

We are mostly interested in $\mu <0$ which suppresses the dark matter density. In theories with antiparticles, the antiparticle distribution involves $-\mu$
which restricts $\vert \mu \vert <m$ \cite{Haber:1981fg,Haber:1981ts}.  This does not apply to the model at hand and the $S$--gas  can be arbitrarily dilute as long as kinetic equilibrium is maintained. At large $ -\mu $,  the density is exponentially suppressed,  $n \propto e^{\mu/T} $.

   %%%%%%%%%%%%%%%%%%%%%%%%%%%%%%%%%%
\begin{figure}[h!]
\begin{center}
 \includegraphics[scale=0.63]{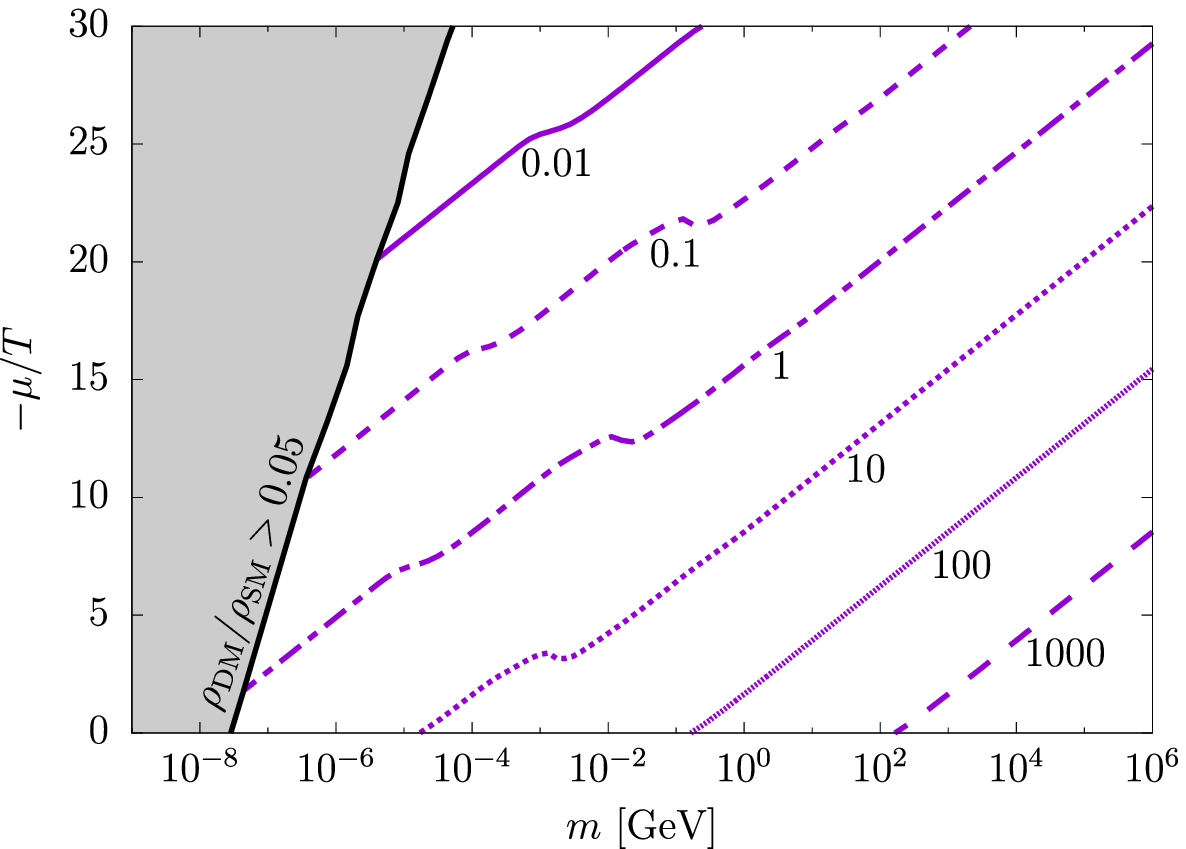}
  \includegraphics[scale=0.63]{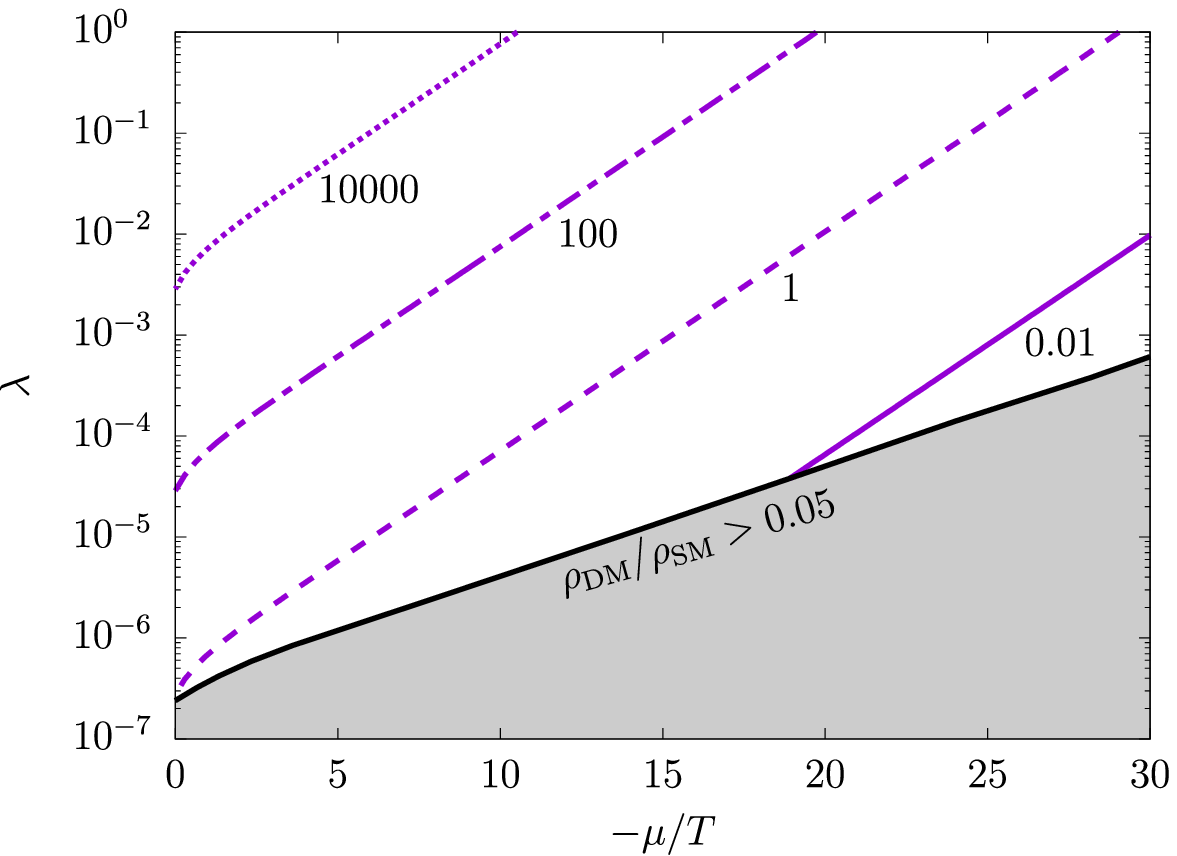}
\end{center}
\caption{   {\it Left:} Parameter space consistent with the correct relic density for dark matter in kinetic equilibrium.   
The curves are labeled by $\xi=T_{\rm SM}/T$. Both  $\xi$ and $\mu/T$ are fixed 
 in the relativistic regime ($T=10m$). In the excluded area, the energy density of relativistic DM exceeds 5\% of that of the SM radiation. 
 {\it Right:}  Minimal self--coupling required for kinetic equilibrium
 as a function of $\mu/T$ for $m=1$ GeV. 
\label{fig:non-thermal}}
\end{figure}
%%%%%%%%%%%%%%%%%%%%%%%%%%%%%%%%%%

It is straightforward to derive the temperature and chemical potential evolution. Since both the total number and entropy are conserved, 
away from the thresholds where the number of    SM degrees of freedom changes,  we have 
\begin{equation}
{n\over T_{\rm SM}^3} =const ~~~,~~~ {s\over T_{\rm SM}^3 }=const \;,
\end{equation}
such that at  $T \gg m$ one finds
\begin{equation}
\mu \propto T \propto T_{\rm SM} \;.
\end{equation}

Our numerical results are shown in Fig.\;\ref{fig:non-thermal}. It displays the constant $\xi$ curves producing the correct DM relic density 
in the $(m, -\mu/T)$ plane.\footnote{We focus on $\mu<0$ since   positive values of $\mu$ are bounded by $m$ and omit the possibility of the Bose--Einstein condensate formation.}  Here the parameters $\xi$ and $\mu/T$ are defined in the relativistic regime ($T=10m$).  
The fixed relic density lines correspond approximately to $e^{-\mu/T} \propto m/\xi^3$. For a fixed $\xi$ and small $m$, the dilution factor $e^{-\mu/T} $ becomes
insignificant and the relativistic dark sector starts making a substantial contribution to the energy density of the Universe.
Throughout this paper we assume that the SM thermal bath dominates the energy density balance, hence we exclude the region
$\rho_{\rm DM }/ \rho_{\rm SM}>0.05$, which is also disfavored by cosmological bounds on the effective number of neutrinos \cite{Tanabashi:2018oca}.  

The right panel of Fig.\;\ref{fig:non-thermal} shows that significantly larger couplings are required for kinetic equilibrium as the gas becomes more dilute.
The rate $\Gamma_{2\rightarrow 4}$ is still smaller than $\Gamma_{2\rightarrow 2}$, so full thermal equilibrium is not reached for a significant range of the couplings.  
We find that the Bullet Cluster bound on the self--coupling 
is  insignificant here and superseded by the constraint $\rho_{\rm DM }/ \rho_{\rm SM}<0.05$.

Note that the dark sector in kinetic equilibrium is allowed to  be significantly hotter than the observable one, $T\gg T_{\rm SM}$. This is consistent with the constraints  due to the exponential suppression of $\rho_{\rm DM }$ for substantial $-\mu/T$. 

Overall, we find that there is a large range of  DM masses consistent with observations. 
For a given $m$ within this range, there exists an initial DM density   which leads to the correct relic abundance. 
Thermodynamical considerations apply to this system as long as the self--coupling is large enough to bring it into kinetic equilibrium.
For smaller couplings, the correct relic abundance can still be obtained, however the dark temperature is ill--defined.
 
\section{Conclusion}

We have performed a comprehensive study of real scalar dark matter decoupled from the 
Standard Model fields. We have considered two regimes where the dark sector can be assigned a temperature: 
\begin{itemize}
\item DM in thermal equilibrium (larger self--coupling)
\item DM in kinetic equilibrium (smaller self--coupling)
\end{itemize}
 In the latter case, the relic abundance is fixed by the initial number density which corresponds to a non--zero effective chemical potential.
 In the former case, it is determined by the freeze--out temperature below which the number changing interactions are suppressed.
  
  We have developed a relativistic approach to the dark matter evolution. In particular, we use fully relativistic expressions for the number changing 
  and number conserving reaction rates. This allows us to explore the relativistic freeze--out regime, which occurs commonly in 
   the allowed parameter space.  
   When  the dark temperature is much smaller than the observable one, the correct DM relic abundance can be obtained for a wide range of DM masses, 10 keV to 100 TeV. The required self--coupling is above $10^{-5}$.
   
  If dark matter reaches only kinetic equilibrium, the correct relic density can be obtained both for $T< T_{\rm SM}$ and $T> T_{\rm SM}$
  as long as DM is sufficiently dilute, 
  $-\mu \gtrsim T$. 
  The allowed DM mass is then above 100 eV.  The presence of chemical potential also suppresses the effect of relativistic DM on nucleosynthesis. 
 
  Altogether, there is vast parameter space consistent with the thermal history of the Universe, while dark matter can be warm or cold.

  \section*{Acknowledgements}

O.L. is indebted to Aleksi Vuorinen for numerous enlightening discussions,  to Keijo Kajantie, Kirill Boguslavski and Howie Haber  for useful communications, and to Eliza Dickie
for verifying some of the analytical results. G.A. thanks Stefan Vogl for  fruitful discussions.
The work of S.P. is partially supported by the National Science Centre, Poland, under research grants DEC-2015/18/M/ST2/00054 and DEC-2016/23/G/ST2/04301.
T.T. acknowledges funding from the Natural Sciences and
Engineering Research Council of Canada (NSERC). 
This research was enabled in part by support provided by Compute Ontario,
WestGrid, Compute Canada, the T30/CIP-cluster at Technical University of
Munich and the Yukawa Institute Computer Facility.

 \section*{Appendix A}

Here we provide some of the explicit expressions for the conversion from a general reference frame to the center--of--mass frame.
Given 2 momenta $p_1,p_2$, we define $p=(p_1+p_2)/2$ and $k=(p_1-p_2)/2$.  
The center--of--mass frame is defined by the relation
\begin{equation}
p=\Lambda(p)~\left( 
\begin{matrix}
&E& \\
&0&\\
&0&\\
&0&
\end{matrix}
\right),
\end{equation}
where $E$ is the particle energy in the center--of--mass frame.
In the convention $p=(p^0, p^3, p^2, p^1)^T$,
the explicit form of $\Lambda$ and its inverse in terms of the rapidity $\eta$ and angular variables $\theta,\phi$ is given by
\begin{eqnarray}
\Lambda(p)=\left( 
\begin{matrix}
1 & 0& 0& 0 \\
0 & 1& 0& 0  \\
0 & 0& \cos\phi & -\sin\phi  \\
0& 0& \sin\phi & \cos\phi 
\end{matrix}
\right)
\left( 
\begin{matrix}
1 & 0& 0& 0 \\
0 & \cos\theta & -\sin\theta & 0  \\
0 & \sin\theta & \cos\theta & 0  \\
0& 0& 0 & 1 
\end{matrix}
\right)
\left( 
\begin{matrix}
\cosh \eta & \sinh \eta& 0& 0 \\
\sinh \eta & \cosh \eta & 0 & 0  \\
0 & 0 & 1 & 0  \\
0& 0& 0 & 1 
\end{matrix}
\right),\nonumber \\
 \Lambda(p)^{-1}= 
\left( 
\begin{matrix}
\cosh \eta & -\sinh \eta& 0& 0 \\
-\sinh \eta & \cosh \eta & 0 & 0  \\
0 & 0 & 1 & 0  \\
0& 0& 0 & 1 
\end{matrix}
\right) 
\left( 
\begin{matrix}
1 & 0& 0& 0 \\
0 & \cos\theta & \sin\theta & 0  \\
0 & -\sin\theta & \cos\theta & 0  \\
0& 0& 0 & 1 
\end{matrix}
\right)
\left( 
\begin{matrix}
1 & 0& 0& 0 \\
0 & 1& 0& 0  \\
0 & 0& \cos\phi & \sin\phi  \\
0& 0& -\sin\phi & \cos\phi 
\end{matrix}
\right). \nonumber
\end{eqnarray}
In this frame, the 6 degrees of freedom of $p_1,p_2$ become $E,\eta$ and 4 angles $\theta,\phi, \theta_k,\phi_k$, where
$\theta_k$ and $\phi_k$ are the spherical coordinate  angles parameterizing $k$. Note that $k^0=0$ and $\vert {\bf{k}} \vert= \sqrt{E^2-m^2}$
due to the  on--shell condition for the initial particles.

As a result, for the rest frame velocity $u=(1,0,0,0)^T$ and final state momenta $k_i$ in the center--of--mass frame, we have 
\begin{equation}
(\Lambda^{-1}u) \cdot k_i = k_i^0 \cosh \eta + k_i^3 \sinh \eta \;,
\end{equation}
which appears in the Bose--Einstein enhancement factors for the final state.
In  the initial state  thermal averaging, we encounter 
\begin{eqnarray}
&&u\cdot p_1= (\Lambda^{-1}u) \cdot (p+k)= E\cosh\eta + \sqrt{E^2 -m^2} \sinh\eta  ~\cos\theta_k \;,\nonumber\\
&&u\cdot p_2= (\Lambda^{-1}u) \cdot (p-k)= E\cosh\eta - \sqrt{E^2 -m^2} \sinh\eta  ~\cos\theta_k \;,
\end{eqnarray}
where we have used $k^3= \vert {\bf k}\vert \cos \theta_k$. 

\section*{Appendix B}

\begin{figure}[h!]
\begin{center}
 \includegraphics[scale=0.63]{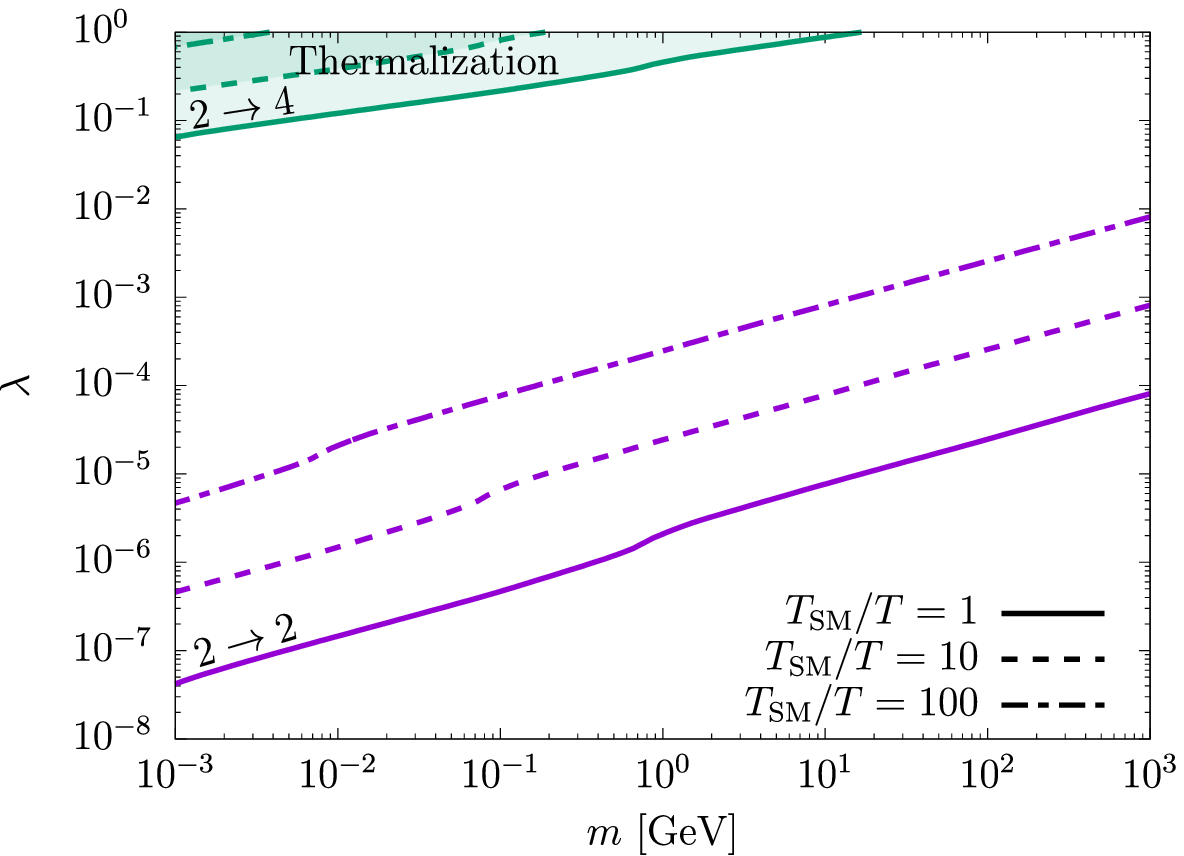}
  \includegraphics[scale=0.63]{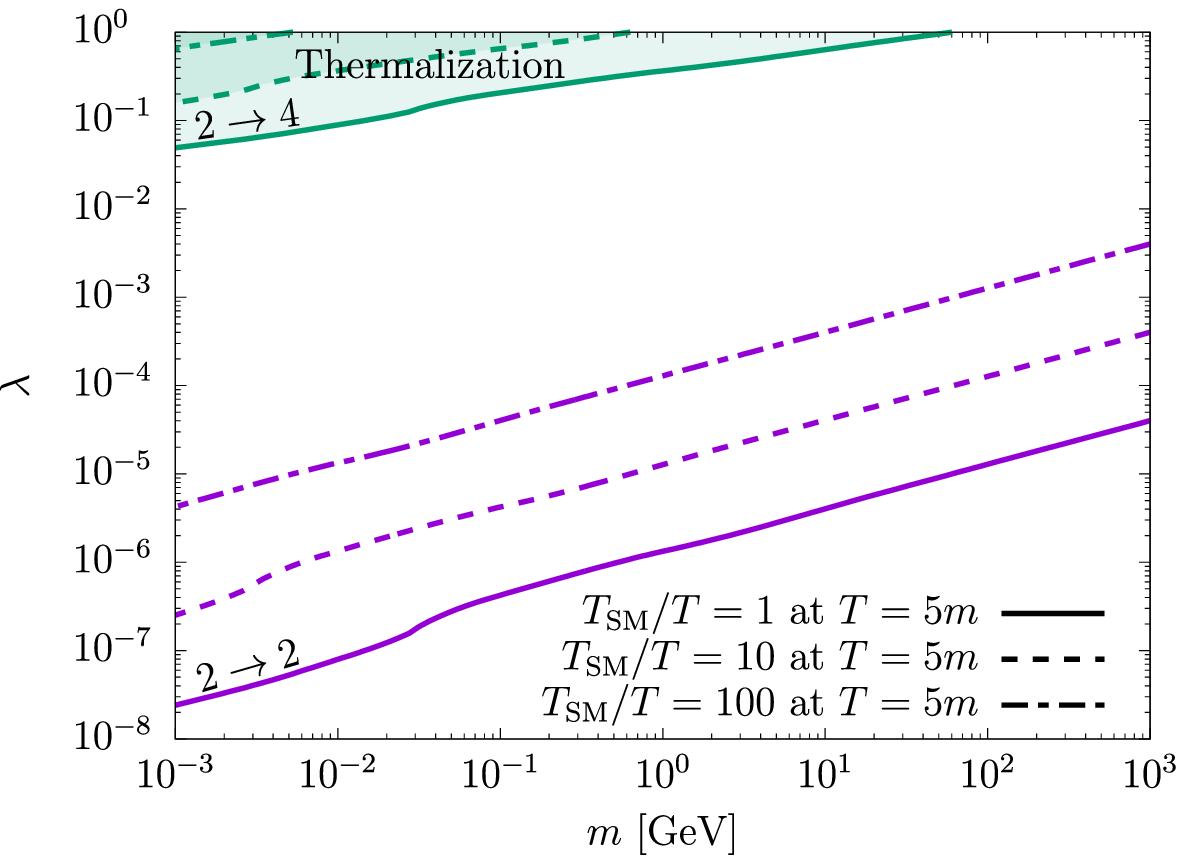}
\end{center}
\caption{   Couplings consistent with thermalization in the non--relativistic regime. The temperature ratio is fixed at $T_0=m/5$ (left)
and $T_0 =5m $ (right).
\label{fig:therm-NR}}
\end{figure}

Here we present a non-relativistic analog of Fig.\;\ref{fig:thermalization}. 
The reaction rate at $\mu \sim 0$ is 
\begin{equation}
\Gamma_{4\rightarrow 2} = 
     \langle    \sigma_{4\rightarrow 2}v^3  \rangle  n_{\rm eq}^4 \;,
   \end{equation}
with a non--relativistic number density $n_{\rm eq}$ and $\langle    \sigma_{4\rightarrow 2}v^3  \rangle$  given by (\ref{NRsigma}).
We require  $T<m/5$ for DM to be sufficiently non--relativistic. 
The condition $3nH<2 \Gamma_{42}$   translates into the lower bound on $\lambda$
shown in Fig.\;\ref{fig:therm-NR}.  The result depends on how $T_{\rm SM}/T$ is fixed: the left panel shows the bounds for  
$T_{\rm SM}/T=1..100$ at $T=m/5$, whereas the right panel displays the bounds for $T_{\rm SM}/T=1..100$ 
fixed in the relativistic regime, 
 $T=5m$, and continued to lower $T< m/5$ 
 using entropy conservation,
\begin{equation}
s/s_{\rm SM} = const \;.
\end{equation}
The reaction rates involve at least $n^2$ and thus drop sharply with the temperature, 
as illustrated in Fig.~\ref{fig:NR}. Therefore, the equilibrium condition for a given $(\lambda,m)$ should be tested at the highest $T$ 
consistent with the non--relativistic regime. If it is not satisfied at this point, dark matter will not thermalize at lower $T$ either.
Similar considerations apply to the kinetic equilibrium condition.

 %%%%%%%%%%%%%%%%%%%%%%%%%%%%%%%%%%
%%%%%%%%%%%%%%%%%%%%%%%%%%%%%%%%%%

The resulting bounds on the coupling are stronger than those in the relativistic case. In particular, comparison of the right panel of Fig.\;\ref{fig:therm-NR} 
with Fig.~\ref{fig:thermalization} shows  that
 if DM is relativistic initially and does not thermalize in the
relativistic regime, it will not reach thermal equilibrium later either.

\section*{Appendix C}

In this appendix, we provide details of the $4 \rightarrow 2$ cross section calculation in the non--relativistic limit. We believe this is useful since  there are significant discrepancies with the results in the literature \cite{Bernal:2015xba,Fairbairn:2018bsw}. 
 
 There are two distinct contributions to the amplitude shown in   Fig.\;\ref{fig:diagram}. 
 The initial state momenta are all $(m, \vec{0})^T$, while the final state momenta can be chosen as $(2m, \sqrt{3}m,0,0)^T$ and
 $(2m, -\sqrt{3}m,0,0)^T$.
  Consider the diagram on the right. The momentum flow through the propagator is $p^2=-3m^2$, while the symmetry factor is $1/(2!) \times12^2 \times 2!4!$, where 
  $1/(2!)$  is due to the 2d order in the coupling and $2!4!$ is due to the permutations among the initial and final state legs.
 For the left diagram, the momentum flow through the propagator is $p^2=9m^2$ and the symmetry factor is 
 $ 1/(2!) \times 96 \times 2!4! $ Thus, the QFT amplitude defined in the $standard$ way is given by 
 \begin{equation}
 \vert  \hat {\cal M}_{4\rightarrow 2} \vert = {1\over 2!} \left(  {\lambda\over 4!}  \right)^2 \left\vert   - {12^2\over 4m^2 } + {96\over 8m^2}      \right\vert  2!4!= {\lambda^2 \over m^2} \;.
\end{equation}
 In our convention, we include the  phase space symmetry factor  for the initial and final state, $1/(2!4!)$, directly in the cross section. Thus, 
 according to (\ref{sigmav3}) we have 
 \begin{equation}
      \langle  \sigma v^3  \rangle  =  \sigma v^3    = {1\over 2! 4!} \times {\sqrt{3} \over 256 \pi m^4}  \;
   \vert  \hat  {\cal M}_{4\rightarrow 2} \vert^2 \;,
   \end{equation}
which reproduces (\ref{NRsigma}).

This result can be verified numerically.  The $2\rightarrow 4$ cross section is calculated with CalcHEP, while in equilibrium the $2\rightarrow 4$ and $4\rightarrow 2$
rates are related through $  \langle  \sigma v^3  \rangle    = \langle  \sigma v  \rangle / n_{\rm eq}^2 $, according to our cross section convention. We find excellent agreement with our analytical formula.

{}

\end{document}